\documentclass[twocolumn, times, twocolappendix, resetfootnote]{aastex7}
\usepackage{amsmath}
\usepackage{graphicx}

\received{*}
\revised{*}
\accepted{*}
\submitjournal{ApJ}

\shorttitle{Coronal heating law}
\shortauthors{Kuznetsov et al.}

\begin{document}
\title{Magneto-Thermal Coupling and Coronal Heating in Solar Active Regions Inferred from Microwave Observations}

\correspondingauthor{Alexey A. Kuznetsov}
\email{a\_kuzn@iszf.irk.ru}

\author[0000-0001-8644-8372]{Alexey A. Kuznetsov}
\affiliation{Institute of Solar-Terrestrial Physics, Irkutsk 664033, Russia}
\email{a\_kuzn@iszf.irk.ru}

\author[0000-0001-5557-2100]{Gregory D. Fleishman}
\affiliation{Center for Solar-Terrestrial Research, Physics Department, New Jersey Institute of Technology, Newark, NJ, 07102-1982}
\affiliation{Institut f\"ur Sonnenphysik (KIS), Georges-Köhler-Allee 401 A, D-79110 Freiburg, Germany}
\email{gfleishm@njit.edu}

\author[0000-0003-2846-2453]{Gelu M. Nita}
\affiliation{Center for Solar-Terrestrial Research, Physics Department, New Jersey Institute of Technology, Newark, NJ, 07102-1982}   
\email{gelu.m.nita@njit.edu}

\author[0000-0002-1107-7420]{Sergey A. Anfinogentov}
\affiliation{Institute of Solar-Terrestrial Physics, Irkutsk 664033, Russia}
\email{anfinogentov@iszf.irk.ru}

\begin{abstract}
The solar corona is much hotter than the photosphere and chromosphere, but the physical mechanism responsible for heating the coronal plasma remains unidentified yet. The thermal microwave emission, which is produced in strong magnetic field above sunspots, is a promising but barely exploited tool for studying the coronal magnetic field and plasma. We analyzed the microwave observations of eight solar active regions obtained with the Siberian Radioheliograph in years $2022-2024$ in the frequency range of $6-12$ GHz. We produced synthetic microwave images based on various coronal heating models, and determined the model parameters that provided the best agreement with the observations. The observations and simulations strongly favour either a steady-state (continuous) plasma heating process, or high-frequency heating by small energy release events with a short cadence. The average magnetic field strength in a coronal loop was found to decrease with the loop length, following a scaling law with the most probable index of about $-0.55$. In the majority of cases, the estimated volumetric heating rate was weakly dependent on the magnetic field strength, and decreased with the coronal loop length following a scaling law with the index of about $-2.5$. Among the known theoretical heating mechanisms, the model based on wave transmission or reflection in coronal loops acting as resonance cavities was found to provide the best agreement with the observations. The obtained results did not demonstrate a significant dependence on the emission frequency in the considered range.
\end{abstract}
\keywords{Solar coronal heating (1989) --- Solar coronal radio emission (1993) --- Solar active regions (1974) --- Astronomy data modeling (1859)}

\section{Introduction}
The solar corona is long known to be hot, with a temperature of about one MK in quiet-Sun areas and up to several MK in active regions \citep[see][and references therein]{Arregui2024}. The fact that the active regions are hotter indicates that the coronal heating is intimately related to the magnetic activity in general and involves a conversion of the magnetic to thermal energy. The coronal heating is usually attributed to either dissipation of some (yet unidentified) waves that transfer energy from lower layers of the solar atmosphere into the corona, or dissipation of tangled magnetic field driven by sub-photospheric plasma motions (which is expected to occur in the form of intermittent nanoflares). Nevertheless, despite of a large number of proposed theoretical models \citep[see, e.g., the reviews of][]{Mandrini2000, Klimchuk2015, Cranmer2019}, the exact physical process responsible for heating the coronal plasma remains elusive.

To identify the coronal heating mechanism, we need to establish where and how this heating occurs, and how the heating process is related to local and global properties of the magnetic field and plasma. To date, most of our knowledge about the thermal plasma in the solar corona has been based on the extreme ultraviolet (EUV) and/or soft X-ray (SXR) observations. These observations have provided important constraints on the plausible heating mechanisms \citep[e.g.,][]{UgarteUrra2019}. However, the coronal emission in the EUV and SXR spectral ranges is optically thin and thus the inferred plasma parameters are only partially spatially resolved (i.e., at each image pixel they are integrated along the line of sight). In addition, the EUV and SXR emissions are sensitive to the coronal abundances (which are not always known, especially in active regions), while not explicitly dependent on the magnetic field.

A complementary, less explored source of information about the thermal coronal plasma is its microwave emission. The microwave emission of solar active regions is produced primarily due to the gyroresonance mechanism \citep[see, e.g.,][and references therein]{Lee2007, Alissandrakis2021}, by thermal electrons gyrating in ambient magnetic field. The emission is produced in gyrolayers---surfaces where the emission frequency equals a harmonic of the local electron cyclotron frequency; in the solar corona, the emission from the third-harmonic gyrolayer typically dominates. The shape of the emission sources reflects the shape of the optically thick gyrolayers (projected onto the image plane), while the brightness temperature of the emission corresponds to the plasma temperature at those gyrolayers, i.e., the emission is sensitive to both the plasma and magnetic field \citep[see, e.g., Figure 1 in][]{Fleishman2025}; there is no explicit dependence on the element abundances \citep{Fleishman2021b}. 

Recently, \citet{Fleishman2021a} have analyzed the coronal heating processes using the microwave observations of the Siberian Solar Radio Telescope and Nobeyama Radioheliograph (at the frequencies of 5.7 and 17 GHz, respectively) in  active region (AR) 11520 on 2012-07-12. They discovered that the synthetic images generated from a 3D magnetic model ``dressed'' with a parametric thermal model \citep{Nita2023} are highly sensitive to parameters of this model and, thus, this sensitivity can be used to identify the best thermal model or a set of best models. However, the study of \citet{Fleishman2021a} had several important limitations, some of which were addressed by \citet{Fleishman2025}, who (i) used a new theory and associated computer codes \citep{Fleishman2021b} that explicitly take into account the multi-temperature plasma distribution in each elementary model element instead of using the distribution moments; (ii) performed systematic searches over an extended parameter space; and (iii) explored a number of coronal heating regimes, including both steady-state heating and various models involving nanoflares. Still, both \citet{Fleishman2021a} and \citet{Fleishman2025} considered only one instance of only one AR and used microwave data obtained with two different instruments, which might be prone of systematic errors due to dissimilar calibrations of the instruments. In this work, we remedy these deficiencies by analyzing the diffuse component of the coronal plasma in additional eight solar active regions observed at different frequencies in the $\sim 6-12$ GHz range obtained with the same instrument. This pool of data allows us to perform a statistical study, and to investigate how the coronal heating characteristics vary from one active region to another.

\section{Observations}
\subsection{Instruments}
The microwave images of the active regions were obtained using the recently commissioned Siberian Radioheliograph \citep[SRH,][]{Altyntsev2020}. This instrument consists of three independent antenna arrays for the frequency bands of about $3-6$, $6-12$, and $12-24$ GHz; it provides 2D images of the Sun at multiple selected frequencies (in the frequency scanning mode). The SRH started regular observations in full configuration in December 2023; additionally, a significant amount of observations (irregular, with separate arrays) were performed during the instrument construction and testing since March 2021. In this study, we analyze mostly those test-mode observations (see also Section \ref{Selection}). We selected observations with the middle-frequency SRH array (in the $\sim 6-12$ GHz range), because they a) cover totally the longest time interval, which allows us to analyze more active regions, b) have a sufficiently high spatial resolution of about $12''-24''$ (that is better than with the low-frequency array), and c) cover the typical frequency range of the gyroresonance emission of solar active regions (at higher frequencies, the microwave flux is usually lower). The observations were performed at 16 equidistant frequencies in the $\sim 6-12$ GHz range (although the exact frequency values were slightly different before and after September 2023), with a cadence of about 3 s and the accumulation time of $\sim 0.1$ s at a single frequency. To improve the signal/noise ratio, we also performed additional averaging over 20 s time intervals, i.e., over $5-6$ subsequent images. The flux calibration was performed using the estimated microwave flux from the quiet-Sun regions \citep{Zirin1991}. Due to the array configuration, the best times for imaging observations with the SRH are $\sim 0.5-1.5$ h before/after the local midday ($\sim$ 05:10 UT): in earlier morning / later evening, the instrument beam becomes too elongated, while just near the midday the images may be adversely affected by sidelobes.

To create the 3D models of the active regions (see Section \ref{Computing}), we used the automatic model production pipeline \citep[AMPP,][]{Nita2023} that employs the magnetograms and optical images from the Helioseismic and Magnetic Imager on board the Solar Dynamic Observatory \citep[\textit{SDO}/HMI,][]{Scherrer2012}.

\subsection{Selection of active regions}\label{Selection}
For the analysis, we selected the active regions that were:

a) located near the central solar meridian (to ensure that the SDO/HMI vector magnetograms had a minimal geometrical distortion);

b) with a well resolved bipolar structure, i.e., with two or more distinctive bright sources in the microwave images; this implied that the active regions were sufficiently large and with $\beta$ or $\beta\gamma$ configuration;

c) isolated, i.e., without other active regions nearby (to ensure that other active regions did not affect the coronal magnetic field connectivity in the considered active region);

d) dominant in terms of microwave emission, i.e., either single or brightest microwave sources on the solar disk (to provide the highest possible dynamic range of the microwave images);

e) without flares at the time of observations (to ensure that the microwave emission had no non-thermal component); the flare occurrence was checked using the Heliophysics Events Knowledgebase\footnote{\url{https://www.lmsal.com/isolsearch}};

f) with a high microwave flux, i.e., with the peak microwave brightness temperature $>10$ times above the quiet-Sun level throughout the entire considered frequency range.

\begin{table}
\caption{List of the solar active regions (ARs) analyzed in this study: NOAA AR numbers, times of observations, locations on the solar disk, maximum and minimum microwave fluxes from the active region (from a $300''\times 300''$ area, in the $\sim 6-12$ GHz range), and maximum and minimum peak brightness temperatures ($T_{\mathrm{b\,max}}$) observed by the SRH in the $\sim 6-12$ GHz range.}
\label{TabARlist}
\setlength{\tabcolsep}{3.0pt}
\movetableright=-1.0cm
\begin{tabular}{ccccc}
\hline\hline
AR & Time, UT & Location & Flux, sfu & $T_{\mathrm{b\,max}}$, MK\\
\hline
12924 & 2022-01-09 04:45 & S27W03 & 13.0$-$15.6 & 0.14$-$0.37\\
      & 2022-01-09 05:35 & S27W03 & 12.4$-$14.9 & 0.15$-$0.36\\
			& 2022-01-10 05:25 & S27W16 & 12.4$-$13.9 & 0.11$-$0.38\\
			& 2022-01-10 06:10 & S27W16 & 12.3$-$15.0 & 0.10$-$0.35\\
\hline
12936 & 2022-01-30 04:15 & N23E03 & 32.7$-$35.3 & 1.06$-$1.25\\
      & 2022-01-30 06:25 & N23E01 & 26.7$-$30.6 & 0.96$-$1.21\\
			& 2022-01-31 04:20 & N23W10 & 21.2$-$23.6 & 0.48$-$0.77\\
			& 2022-01-31 06:15 & N23W11 & 22.4$-$24.7 & 0.56$-$0.90\\
\hline
13007 & 2022-05-14 03:25 & S19E04 & 14.8$-$17.4 & 0.20$-$1.76\\
      & 2022-05-14 06:35 & S19E02 & 15.4$-$19.5 & 0.21$-$1.67\\
      & 2022-05-15 03:45 & S19W08 & 14.0$-$16.7 & 0.20$-$1.31\\
      & 2022-05-15 06:20 & S19W09 & 13.9$-$17.7 & 0.18$-$1.28\\
\hline				
13234 & 2023-02-26 03:40 & N31E01 & 16.3$-$21.3 & 0.18$-$0.88\\
      & 2023-02-26 06:40 & N31W00 & 20.5$-$23.7 & 0.33$-$1.19\\
      & 2023-02-27 03:45 & N31W12 & 18.2$-$21.2 & 0.16$-$0.83\\
      & 2023-02-27 06:45 & N31W14 & 21.6$-$26.2 & 0.25$-$1.00\\
\hline				
13245 & 2023-03-09 04:00 & S16E08 & 16.8$-$18.5 & 0.51$-$1.18\\
      & 2023-03-09 06:25 & S16E07 & 17.1$-$18.5 & 0.63$-$1.46\\
      & 2023-03-10 03:45 & S16W04 & 18.2$-$22.6 & 0.80$-$1.42\\
      & 2023-03-10 06:20 & S16W04 & 19.0$-$24.2 & 1.02$-$1.68\\
\hline				
13315 & 2023-05-27 03:50 & S16E04 & 21.8$-$25.0 & 1.17$-$2.08\\
      & 2023-05-27 06:30 & S16E02 & 25.8$-$29.6 & 1.36$-$2.01\\
      & 2023-05-28 03:55 & S16W10 & 21.1$-$27.6 & 0.40$-$1.82\\
      & 2023-05-28 06:50 & S16W11 & 23.2$-$29.2 & 0.56$-$1.86\\
\hline
13372 & 2023-07-17 03:45 & N19E12 & 20.5$-$23.2 & 0.62$-$1.27\\
      & 2023-07-17 06:15 & N19E10 & 20.4$-$23.0 & 0.64$-$1.28\\
      & 2023-07-18 04:00 & N19W00 & 21.4$-$24.8 & 0.84$-$1.74\\
      & 2023-07-18 06:40 & N19W01 & 24.3$-$28.1 & 0.79$-$1.52\\
\hline 				
13559 & 2024-01-23 04:20 & N32E09 & 31.0$-$38.1 & 0.20$-$0.71\\
      & 2024-01-23 05:50 & N32E09 & 27.3$-$37.4 & 0.23$-$0.82\\
      & 2024-01-24 04:15 & N32W04 & 23.2$-$29.3 & 0.23$-$0.91\\
      & 2024-01-24 06:15 & N32W06 & 21.4$-$27.5 & 0.19$-$0.85\\							
\hline
\end{tabular}
\end{table}

The selected active regions (eight in total) are presented in Table \ref{TabARlist}. We note that the above conditions (c-e) were mostly satisfied at the rise phase of the 25th solar cycle; near the peak of the cycle, the solar disk became too cramped with active regions, and the flares became too frequent, so that finding a sufficiently isolated and dominant non-flaring active region became difficult. For this reason, most of the selected active regions (except one) were observed by the SRH in years $2022-2023$, i.e., during the commissioning phase. For each of the selected active regions, we analyzed the observations on two subsequent days, and at two times for each day (typically, before and after the local midday), i.e., four image sets for each active region and 32 image sets in total. We also note that the microwave fluxes in Table \ref{TabARlist} are unambiguous characteristics of the active regions, while the peak brightness temperatures may be dependent on the spatial resolution of the used instrument because of corresponding dilution of unresolved microwave flux. Magnetograms and optical, EUV, and microwave images of the selected active regions are presented in Appendix D in the ApJ version of the paper.

\begin{figure*}
\centerline{\includegraphics{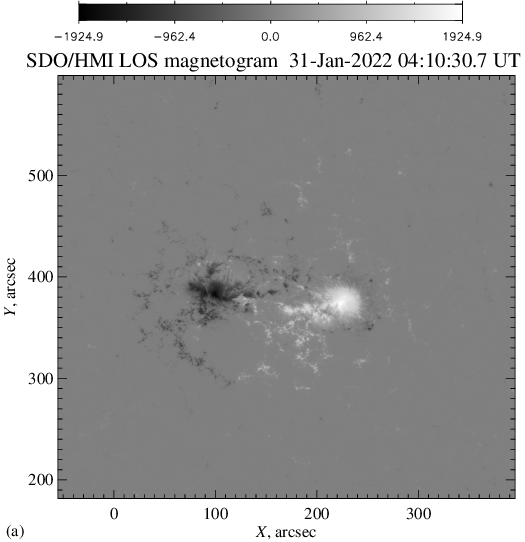}
\includegraphics{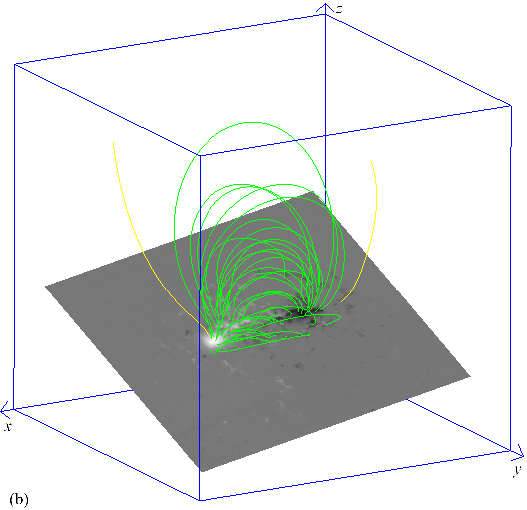}}
\caption{(a) \textit{SDO}/HMI magnetogram of  AR 12936 on 2022-01-31. (b) Model of the active region (screenshot from GX Simulator) with the field lines demonstrating the structure of the coronal magnetic field. The vertical direction ($z$ axis) corresponds to the direction to the Earth. The model was rotated around the line of sight by $\sim 180^{\circ}$ to provide a better view.}
\label{FigARstructure}
\end{figure*}

\section{Methods}
\subsection{Computing the synthetic microwave images}\label{Computing}
Since we are going to compare the microwave observations with the predictions of various coronal heating models, the key part of the method is computing the synthetic microwave images of solar active regions. We extended the approach proposed by \citet{Fleishman2021a}, and employed the interactive modeling package GX Simulator \citep{Nita2018, Nita2023}. For each of the selected active regions and each considered time, we created a ``master'' 3D model of the active region, using the AMPP. Such a model included a 3D magnetic field cube obtained using the \textit{SDO}/HMI vector photospheric magnetogram (at the time closest to the observation time) and the nonlinear force-free extrapolation code \citep[see][]{Nita2023}. A model consisted of the coronal and chromospheric parts (above/below $\sim 2000$ km). The chromospheric part was automatically populated with thermal plasma according to the solar atmospheric models by \citet{Fontenla2009}, see \citet{Nita2018, Nita2023}. The dimensions of the models were chosen to be $290\times 290\times 185.6$ $\textrm{Mm}^3$ to include fully the considered active regions, with the spatial resolution of 1450 km in the coronal part; in the chromospheric part of the models, the height resolution was finer and variable. Figure \ref{FigARstructure} demonstrates an example of the \textit{SDO}/HMI magnetogram (for AR 12936) and the corresponding structure of the coronal magnetic field obtained using the GX Simulator package.

\subsubsection{Thermal plasma model}
The coronal part of an active region model was populated with thermal plasma according to a specified parametric coronal heating model. For each volume element in the corona, we computed the magnetic field line passing through the center of that volume element. Then, for the volume elements associated with closed field lines, the volumetric heating rate $Q$ (measured in erg $\textrm{cm}^{-3}$ $\textrm{s}^{-1}$) was described by the parametric heating law in the form of
\begin{equation}\label{HeatingRate}
Q=Q_0\left(\frac{\left<B\right>}{B_0}\right)^a\left(\frac{L}{L_0}\right)^{-b},
\end{equation}
where $\left<B\right>$ is the magnetic field strength averaged along the field line, $L$ is the length of the field line, $B_0$ and $L_0$ are some normalization constants (chosen to be 100 G and $10^9$ cm, respectively), $Q_0$ characterizes the heating intensity, and the power-law indices $a$ and $b$ depend on the physical heating mechanism. This model assumes heating along individual magnetic flux tubes; since the thermal conductivity of plasma along magnetic field is very high, the exact distribution of the heating sources along a flux tube is not important. We note that the parameter $Q_0$ itself is not representative, because it does not allow a direct comparison between the heating models with different $a$ and/or $b$ values.

The resulting thermal plasma parameters in the above-mentioned volume elements were obtained from hydrodynamical simulations, using the Enthalpy-Based Thermal Evolution of Loops code \citep[EBTEL,][]{Klimchuk2008, Cargill2012a, Cargill2012b, Barnes2016, Schonfeld2020}, with the latest implementation (known as EBTEL++) presented in \citet{Schonfeld2022}. This code simulates the plasma heating/cooling processes, with account for chromospheric evaporation and condensation. For given heating rate $Q$ and magnetic flux tube length $L$, the EBTEL code computes the corresponding equilibrium characteristics of thermal plasma: the differential emission measure (DEM) and differential density metric \citep[DDM,][]{Fleishman2021b}. The GX Simulator package contains several pre-computed tables of the EBTEL code outputs corresponding to different heating regimes (see Section \ref{HeatingRegime}); for a given $(Q, L)$ pair, the corresponding plasma DEM and DDM were computed by interpolation within a chosen EBTEL table. For the volume elements associated with open magnetic field lines (or, more correctly, with the field lines closed beyond the considered data cube, so that we were unable to determine the line length and average magnetic field), or where the $Q$ or $L$ parameters fell beyond the considered EBTEL table, the plasma was described by the isothermal barometric model (with the density of $10^8$ $\textrm{cm}^{-3}$ at the corona base and the temperature of 1 MK).

\subsubsection{Computing the microwave emission}
After determining the magnetic field and plasma distributions within an active region model, the corresponding gyroresonance and free-free microwave emissions were computed according to the refined theory developed by \citet{Fleishman2021b} using the code documented in \citet{Kuznetsov2021}. In particular, the gyroresonance emission was computed using either the plasma DDM in the volume elements where the DDM was available, or the single-temperature plasma density and temperature values otherwise; the free-free emission was computed using either the average plasma density and temperature values derived from the DDM, or the single-temperature plasma parameters. 

\begin{figure*}
\centerline{%
\includegraphics{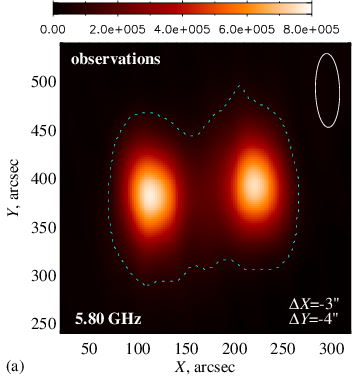}%
\includegraphics{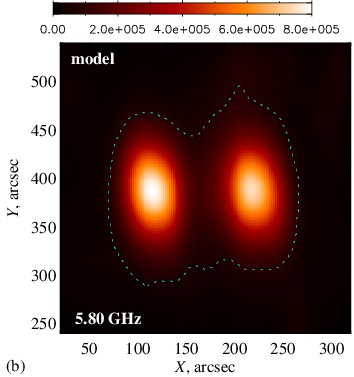}%
\includegraphics{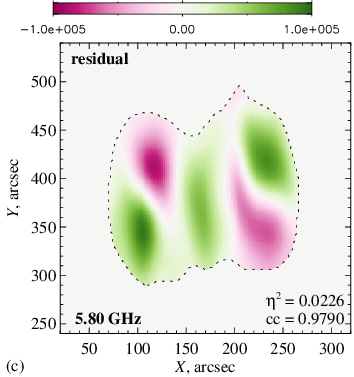}}
\centerline{%
\includegraphics{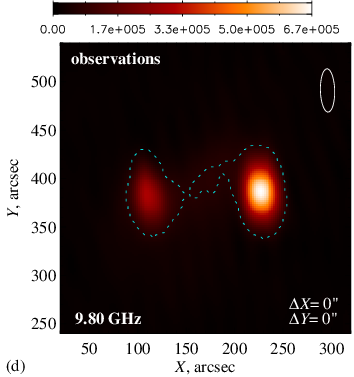}%
\includegraphics{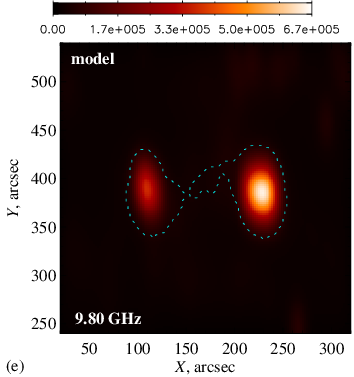}%
\includegraphics{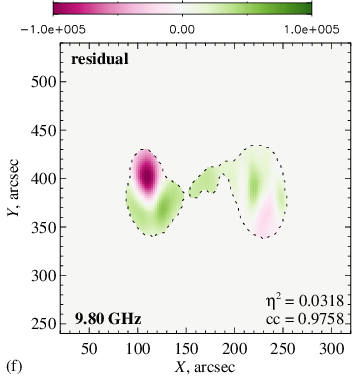}}
\caption{Comparison of the observed (with the SRH) and synthetic (model) microwave images of AR 12396, at 2022-01-31 04:20 UT. Each row demonstrates the maps of the observed ($I_{\mathrm{obs}}$) and synthetic ($I_{\mathrm{mod}}$) brightness temperatures, as well as of the corresponding residual $I_{\mathrm{obs}}-I_{\mathrm{mod}}$. The synthetic images were computed for the steady-state EBTEL heating regime, for $a=0.3$, $b=2.7$, and $Q_0=0.0210$ erg $\textrm{cm}^{-3}$ $\textrm{s}^{-1}$ at the frequency of 5.80 GHz (top row), and for $a=-0.2$, $b=2.5$, and $Q_0=0.0207$ erg $\textrm{cm}^{-3}$ $\textrm{s}^{-1}$ at the frequency of 9.80 GHz (bottom row). The dotted contours show the areas used to compute the fitting quality metrics (the residuals are shown only within those areas). In panels (a) and (d), the white ellipses are the corresponding SRH beam contours at 1/2 level, and $\Delta X$ and $\Delta Y$ are the shifts applied to align the observed and synthetic images. In panels (c) and (f), the corresponding values of the $\eta^2$ metric and correlation coefficients (cc) are shown.}
\label{FigImgComparison}
\end{figure*}

Creating a master model, populating it with thermal plasma, and computing the resulting emission can be performed interactively, using the GX Simulator tool, as has been done in the work of \citet{Fleishman2021a}. To facilitate faster and automatic computations, we have now developed a standalone code\footnote{\url{https://github.com/kuznetsov-radio/gximagecomputing}} for computing the synthetic microwave images in a batch mode; version 1.0.0 is archived in Zenodo \citep{Kuznetsov2025a}. This code uses the master model of an active region (created with the GX Simulator AMPP), an EBTEL output table, and the coronal heating model parameters $Q_0$, $a$, and $b$ as input, and computes the microwave emission maps (in terms of brightness temperature) at the specified frequencies. In this study, we chose for those maps a $300''\times 300''$ (or $360''\times 300''$ for AR 13559) field of view, with the resolution of $2''$. Finally, the obtained microwave emission maps were convolved with the SRH response functions. Examples of the resulting synthetic microwave images at two frequencies (for the same active region as shown in Figure \ref{FigARstructure}) are demonstrated in Figure \ref{FigImgComparison}b,e.

\subsection{Fitting models to observations}
\subsubsection{Quantifying the fitting quality}
To compare quantitatively the observed and synthetic microwave images, the observed images were re-binned and truncated to match the resolution and field of view of the synthetic ones. Also, since interferometric instruments do not always provide accurate positioning, the observed images were additionally shifted (typically, by up to a few arcseconds) to provide the maximum cross-correlation with the synthetic images. The difference between the observed and synthetic images (i.e., the goodness of the model fit) can be characterized by different metrics, e.g.,
\begin{equation}\label{RhoMetric}
\rho^2=\left<\frac{(I_{\mathrm{obs}}-I_{\mathrm{mod}})^2}{I_{\mathrm{obs}}^2}\right>,
\end{equation}
\begin{equation}\label{ChiMetric}
\chi^2=\left<\frac{(I_{\mathrm{obs}}-I_{\mathrm{mod}})^2}{\sigma_{\mathrm{obs}}^2}\right>,
\end{equation}
\begin{equation}\label{EtaMetric}
\eta^2=\frac{\left<(I_{\mathrm{obs}}-I_{\mathrm{mod}})^2\right>}{\left<I_{\mathrm{obs}}\right>^2},
\end{equation}
where $I_{\mathrm{obs}}$ is the observed brightness, $\sigma_{\mathrm{obs}}$ is the uncertainty of the observations, and $I_{\mathrm{mod}}$ is the synthetic (model) brightness; all these values are attributed to individual pixels and are dependent on the image coordinates $x$ and $y$. All averagings in Equations (\ref{RhoMetric}--\ref{EtaMetric}) were performed over the area where either observed or synthetic microwave brightnesses were sufficiently high to satisfy the condition 
\begin{displaymath}
I_{\mathrm{obs}}>\alpha\max(I_{\mathrm{obs}})\quad\textrm{or}\quad I_{\mathrm{mod}}>\alpha\max(I_{\mathrm{mod}});
\end{displaymath} 
here, we used the threshold value of $\alpha=0.1$.

\subsubsection{The fitting procedure}
\begin{figure*}
\centerline{\large $\eta^2$ metric}\vspace{3pt}
\centerline{%
\includegraphics{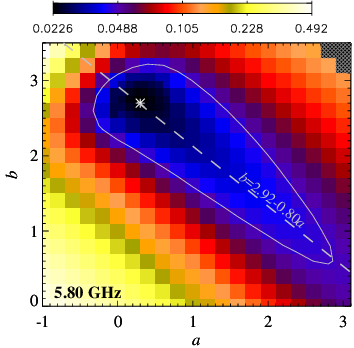}%
\includegraphics{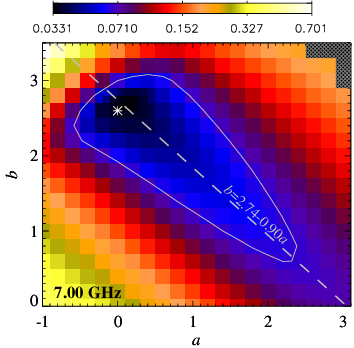}%
\includegraphics{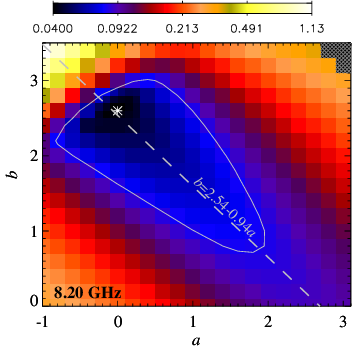}}
\centerline{%
\includegraphics{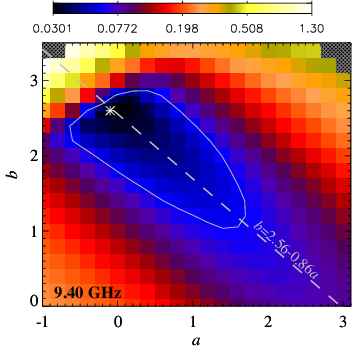}%
\includegraphics{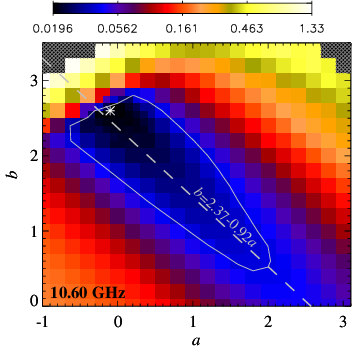}%
\includegraphics{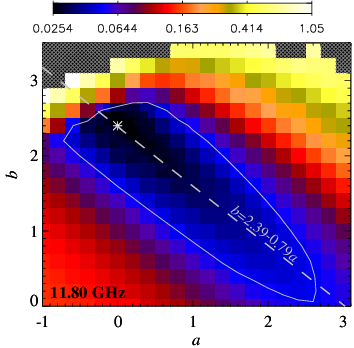}}
\caption{Maps of the fitting quality metric $\eta^2$ vs. the heating law parameters $(a, b)$, for AR 12396, at 2022-01-31 04:20 UT, for the steady-state EBTEL heating regime, at six selected frequencies. The cross-hatched areas are those where the local one-dimensional minimization procedure (over the $Q_0$ parameter) failed, due to the limited EBTEL output table. The minima of $\eta^2$ are marked by asterisks, solid white lines represent the contours of $\eta^2=2\eta^2_{\min}$, and dashed lines represent linear fits to the best-fit (``degeneracy'') stripes in the $(a, b)$ space; the equations of those linear fits are shown, too.}
\label{FigMultiFreqs}
\end{figure*}

For a given active region at given time and frequency, with a selected EBTEL heating regime, the task of finding the best-fit coronal heating model is reduced to minimizing a chosen fitting quality metric ($\rho^2$, $\chi^2$, or $\eta^2$) as a function of three parameters: $Q_0$, $a$, and $b$, see Equation (\ref{HeatingRate}). The minimization was performed using the Coronal Heating Modeling Pipeline\footnote{\url{https://github.com/kuznetsov-radio/gxmodelfitting}} (CHMP) code, version 1.0.0 of which is archived in Zenodo \citep{Kuznetsov2025b}. This code uses an observed microwave image (from the SRH or another instrument), a master GX Simulator model of an active region, and an EBTEL lookup table as input data. For each pair of power-law indices $a$ and $b$ from a specified grid, the CHMP code computes the heating constant $Q_0$ that minimizes the chosen fitting quality metric, thus providing a 2D map of that metric vs. $a$ and $b$, hereafter referred to as $a{-}b$ diagram (see, e.g., Figure \ref{FigMultiFreqs}). Such $a{-}b$ diagrams can then be used to find the best-fit $(a, b)$ combination that provides the global minimum of the metric, to estimate the uncertainties, and to explore other characteristics of the obtained solution. The CHMP code also provides an opportunity to find automatically a local minimum of the chosen fitting quality metric in the 3D parameter space $(Q_0, a, b)$, and (optionally) to explore an area around that minimum in the $(a, b)$ space, where the fitting quality metric is below the certain threshold. All minimizations are performed separately at each frequency, because the solutions at different frequencies can be different as well (see Section \ref{FrequencyDependence}); at the same time, multi-frequency observations allow for some speed optimization. We also note that for any given $(a, b)$ combination, minimization over $Q_0$ is performed within a certain range limited by the chosen EBTEL output table: the number of volume elements where the parameters of the associated closed magnetic field line fall beyond the EBTEL table should not exceed 10\% of the total number of volume elements associated with closed magnetic field lines.

The $\rho^2$ metric (\ref{RhoMetric}) was considered earlier, e.g., in the work of \citet{Fleishman2021a}; however, this metric tends to overestimate the contribution of the regions with a low observed brightness $I_{\mathrm{obs}}$. The ``classical'' $\chi^2$ metric (\ref{ChiMetric}) has a well understood statistical meaning; however, it requires knowing the measurement uncertainty $\sigma_{\mathrm{obs}}$ at individual pixels (not reduced to the general calibration uncertainty), which proved to be difficult to obtain for the SRH observations. The $\eta^2$ metric (\ref{EtaMetric}) has been found to be the optimal one: it is stable, i.e., small variations of the model parameters result in comparably small changes of $\eta^2$; in addition, the minimum values of $\eta^2$ have been found to correspond to the best visual agreement between the observed and synthetic images (cf. Appendix \ref{MultiImageComparison}). Therefore, we focused on the $\eta^2$ metric in this study.

In this study, for each model setup, the $(a, b)$ parameter space was firstly explored with a grid size of 0.2 in both directions, to find roughly the minimum of the $\eta^2$ metric and investigate the general shape of the $\eta^2(a, b)$ distribution. Then, to refine the results, an area around that minimum (namely, where $\eta^2<1.1\eta^2_{\min}$) was further explored with a grid size of 0.1 in both directions. The computations were performed using the Seismocorona cluster at the Institute of Solar-Terrestrial Physics.

\section{Results}\label{Results}
An example of fitting the coronal heating model to the SRH observations (for AR 12396, at the frequencies of 5.80 and 9.80 GHz) is shown in Figure \ref{FigImgComparison}. The heating model parameters $a$, $b$, and $Q_0$ were chosen to provide the best fits, i.e., the minimum values of the $\eta^2$ metric. One can see that the above-described method indeed allows us to reproduce the observed microwave emission of solar active regions with a high accuracy, which is confirmed by both a good visual agreement between the observed and synthetic images, and reasonably low values of the fitting quality metric ($\eta^2\simeq 2-3\%$). More synthetic microwave images for different $(a, b)$ values (including those that do not provide the best fit) are presented in Figure \ref{MultiABcomparison} in Appendix \ref{MultiImageComparison}.

We estimated the measurement uncertainty in the SRH microwave images following the approach outlined by \citet{Bastian2025a, Bastian2025b}. For the considered here images of active regions, the relative uncertainty $\sigma_I/I_{\mathrm{obs}}$ varied from $\sim 0.3-0.5\%$ at the intensity peak to $\sim 3-5\%$ in the regions with ten times lower intensity. At the same time, as follows from Figure \ref{FigImgComparison}, the remaining residuals are considerably larger---from $\sim 1-2\%$ near the intensity peaks up to more than 50\% at the edges of the considered area, i.e., the uncertainty of the model dominates. This implies that even the best-fit model is oversimplified as has been outlined by \citet{Fleishman2025}. Solar active regions may contain isolated narrow coronal loops where the heating regime or/and intensity is different from those in the diffuse component of the active region \citep{Fleishman2025}. ``Selectively heated'' loops are not yet reproduced by our thermal models; in addition, some of the needed connectivity might not be present in the employed magnetic model, which itself may need further improvement. In the future, we are going to use these small but statistically significant mismatches to further improve our magneto-thermal modeling. Nevertheless, we have found (see Appendix \ref{MultiImageComparison}) that the differences between heating models with different $a$ and $b$ indices or/and heating regimes  are meaningful as they well exceed the measurement uncertainties in the microwave images. Thus, by minimizing the overall fitting quality metric, we can identify the most favourable diffuse coronal heating model among the considered range of models.

\subsection{General structure of the solutions}
Figure \ref{FigMultiFreqs} demonstrates a typical example of $a{-}b$ diagrams, or maps representing the dependences of the model-to-observations fitting quality metric $\eta^2$ on the heating law parameters $a$ and $b$, at six selected frequencies. The diagrams were obtained for AR 12396, at 2022-01-31 04:20 UT, assuming the steady-state EBTEL heating regime (this choice will be explained in Section \ref{HeatingRegime}). One can see that the best agreement between the model and observations is achieved at $a\simeq 0.0$ and $b\simeq 2.6$, and these best-fit parameters are quite similar at different frequencies; the best-fit $\eta^2$ metric values are of about $2-4\%$.

Meanwhile, the model parameters providing a comparably good agreement with observations (i.e., with low $\eta^2$, which is indicated by black-blue shades in the diagrams) form rather long and narrow stripes in the $(a, b)$ space. This occurs because the parameters of the magnetic flux tube $\left<B\right>$ and $L$ in the heating law (\ref{HeatingRate}) are correlated with each other, namely, longer flux tubes tend to have weaker average magnetic fields. The relationship between the magnetic flux tube parameters can be  approximated by a scaling law in the form of \citep{Klimchuk1995}
\begin{equation}\label{Bscale}
\left<B\right>\approx B'_0\left(\frac{L}{L_0}\right)^{\delta},
\end{equation}
where $B'_0$ is an additional normalization constant (not necessarily the same as in Equation (\ref{HeatingRate})), and $\delta$ is a constant power-law index (typically, negative). In this case,  parametric heating law (\ref{HeatingRate})  reduces to (\citealt{Kuznetsov2022}; T. Kucera et al. 2025, in preparation)
\begin{equation}\label{HeatingRate1}
Q\approx Q_0\left(\frac{B'_0}{B_0}\right)^a\left(\frac{L}{L_0}\right)^{a\delta-b}=Q'_0\left(\frac{L}{L_0}\right)^{a\delta-b},
\end{equation}
where the factor $Q'_0$, like $Q_0$, has no direct physical meaning, and can be determined for each $(a, b)$ combination by minimizing a chosen metric. Thus the heating rate becomes approximately dependent on the magnetic flux tube length only, and the heating model becomes degenerate with respect to the parameters $a$ and $b$: with an appropriate (i.e., optimal in terms of model-to-observations fitting) choice of the factor $Q'_0$, any combination of $a$ and $b$ satisfying the condition 
\begin{equation}\label{Degeneration}
a\delta-b=\textrm{const}
\end{equation} 
will provide exactly the same heating rates and hence the same synthetic observables, such as the microwave emission maps. Relationships (\ref{Degeneration}) between the heating model parameters are directly reflected in the $a{-}b$ diagrams as the characteristic stripes of low values of the fitting quality metric (``degeneracy stripes''). Thus the $a{-}b$ diagrams can be used to diagnose the scalings between the parameters of magnetic flux tubes in solar active regions, see Section \ref{delta_b0_scalings}. To obtain quantitative parameters of the degeneracy stripes, we fitted them with straight lines in the form of $b=b_0+a\delta$, as described in Appendix \ref{StripeFitting}; the corresponding fits are shown in Figure \ref{FigMultiFreqs} as well. These linear fits are discussed in the next Sections in more detail.

For the $(a, b)$ combinations providing the best model-to-observations agreement, i.e., located at the axis of the degeneracy stripe, the heating law (\ref{HeatingRate1}) becomes
\begin{equation}\label{HeatingRate2}
Q=Q'_0\left(\frac{L}{L_0}\right)^{-b_0}.
\end{equation}
Thus the intercept $b_0$ of the degeneracy stripes characterizes the effective dependence of the heating rate on the magnetic flux tube length, including both direct and indirect (via the scaling law for the magnetic field strength) dependences. In the paper of \citet{Klimchuk1995}, this dependence was characterized by the power-law index $\alpha$, with $\alpha=-b_0$.

Theoretically, if Equations (\ref{HeatingRate}) and (\ref{Bscale}--\ref{HeatingRate1}) described the structure of and processes in the active regions exactly, the degeneracy stripes in the $a{-}b$ diagrams would be infinitely long, and the fitting quality metrics would depend only on the coordinate across a respective degeneracy stripe, while not on the coordinate along that stripe. Actually, as seen in Figure \ref{FigMultiFreqs}, the stripes have finite lengths, with abrupt edges at the upper-left ends (at small $a$ and large $b$ values), and a gradual increase of the fitting quality metric along the stripes in the lower-right direction (towards larger $a$ and smaller $b$ values). These variations along the degeneracy stripe likely occur because scaling law (\ref{Bscale}) is not a functional dependence but rather a regression between widely scattered data points, while the characteristics of individual flux tubes can substantially deviate from it \citep[see, e.g., Figure 3a in the paper of][]{Fleishman2021a}. 

The described degeneracy of the heating model implies that we are unable to determine reliably a single unique best-fit combination of the heating model parameters. Instead, the presented simulations can be used to identify the subsets of the $(a, b)$ parameters that are consistent (or not consistent) with observations, which imposes certain constrains on the plausible coronal heating mechanisms, as will be discussed in Section \ref{HeatingMechanisms}.

\begin{figure}
\centerline{\includegraphics{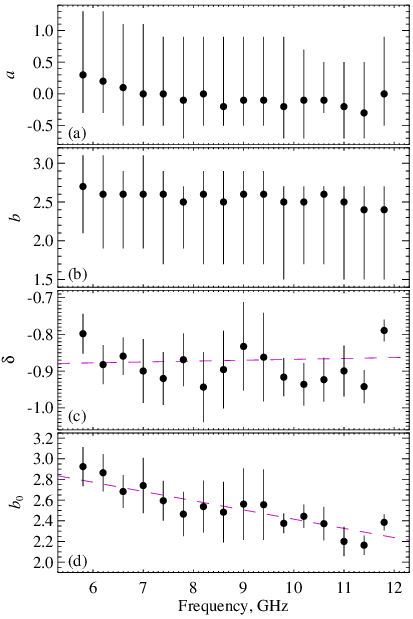}}
\caption{Best-fit parameters of the coronal heating model for AR 12396, at 2022-01-31 04:20 UT, for the steady-state EBTEL heating regime, at different frequencies. (a--b) The best-fit values of the power-law indices $a$ and $b$; the vertical lines mark the respective ranges where $\eta^2<2\eta^2_{\min}$. (c-d) The slopes $\delta$ and intercepts $b_0$ of the best-fit (``degeneracy'') stripes in the $a{-}b$ diagrams; the vertical lines mark the $1\sigma$ uncertainties. The dashed lines represent the linear fits.}
\label{FigFitParms}
\end{figure}

\subsection{Dependence on the emission frequency}\label{FrequencyDependence}
As has been said in the previous Section, the characteristics of the coronal heating model inferred from the microwave images at different frequencies are similar. Figure \ref{FigFitParms} demonstrates the best-fit values of the power-law indices $a$ and $b$, as well as the characteristics of the degeneracy stripes $\delta$ and $b_0$, for the same active region as above (AR 12396 at 2022-01-31 04:20 UT), at all 16 available SRH frequencies. One can see that the best-fit parameters $a$ and $b$ at different frequencies are indeed nearly the same (with $\left<a\right>\simeq 0.0$ and $\left<b\right>\simeq 2.6$), without noticeable frequency trends. The slope $\delta$ of the degeneracy stripes demonstrates no significant variations with frequency (the variations are irregular and small in comparison with the uncertainties in determining that parameter); the intercept $b_0$ tends to decrease slightly with frequency, i.e., the degeneracy stripes shift downwards on the $a{-}b$ diagrams. Since the degeneracy stripes at different frequencies are nearly parallel, their intersection point (that is expected to correspond to the ``true'' unique combination of $a$ and $b$) is poorly defined, i.e., the uncertainties in determining that point are too large ($\gg 1$). Perhaps, having a broader frequency range would help to remove this degeneracy.

\begin{figure}
\centerline{\includegraphics{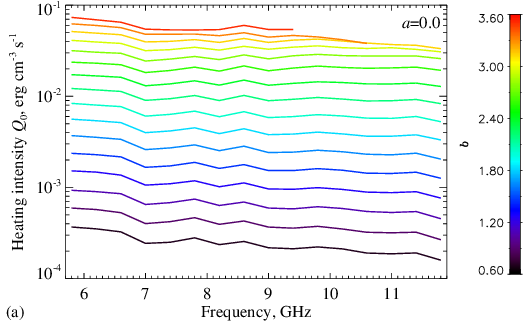}}
\centerline{\includegraphics{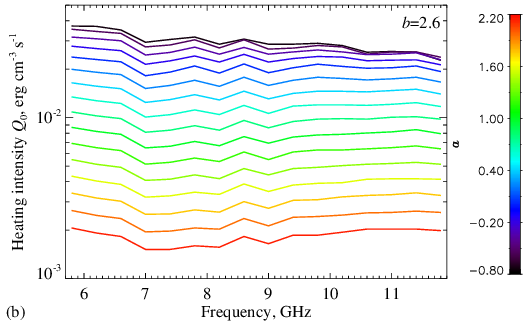}}
\caption{Best-fit heating intensities $Q_0$ for AR 12396, at 2022-01-31 04:20 UT, for the steady-state EBTEL heating regime, at different frequencies: (a) for $a=0.0$ and different values of $b$, and (b) for $b=2.6$ and different values of $a$.}
\label{FiqQ0profiles}
\end{figure}

Figure \ref{FiqQ0profiles} demonstrates the frequency dependences of the best-fit heating intensity $Q_0$ for fixed combinations of $a$ and $b$ indices. We remind that the factor $Q_0$ in Equation (\ref{HeatingRate}) has no straightforward physical meaning, and a direct comparison of the $Q_0$ values for different $(a, b)$ combinations is meaningless; nevertheless, we can compare the $Q_0$ values for the same $(a, b)$ but different emission frequencies. One can see from the figure that the best-fit $Q_0$ values at different frequencies are  comparable, and the weakest frequency variations of $Q_0$ correspond to the $a$  and $b$ values that provide the best model-to-observations fitting quality (in terms of the lowest $\eta^2$ metric) as well, i.e., $a\simeq 0.0$, $b\simeq 2.6$. For the $(a, b)$ combinations significantly different from the mentioned one, the frequency variations of the best-fit $Q_0$ value become more pronounced.

\begin{figure}
\centerline{\includegraphics{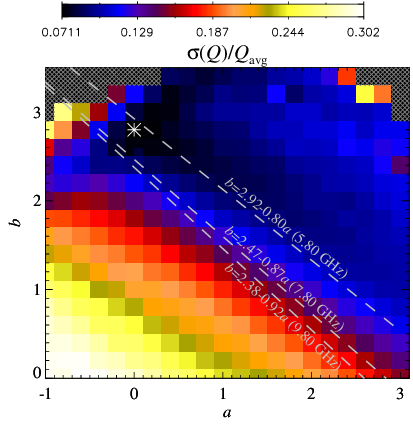}}
\caption{The relative dispersion of the best-fit heating rate $Q_0$ over frequency, vs. the $a$ and $b$ parameters, for AR 12396, at 2022-01-31 04:20 UT, for the steady-state EBTEL heating regime. The minimum of $\sigma(Q_0)/\left<Q_0\right>$ is marked by an asterisk. The linear fits to the best-fit (``degeneracy'') stripes in the $(a, b)$ space (same as in Figure \protect\ref{FigMultiFreqs}) at three selected frequencies are shown by dashed lines; the equations of those linear fits are shown, too.}
\label{FigSigmaQ0}
\end{figure}

Figure \ref{FigSigmaQ0} demonstrates the 2D map (vs. $a$ and $b$) of a quantitative characteristic of the variations of the best-fit $Q_0$ parameter with frequency: the normalized standard deviation $\sigma(Q_0)/\left<Q_0\right>$; this value is shown only at the $(a, b)$ points where the fitting procedure was successful at all 16 considered frequencies. The minimum variations, with $\sigma(Q_0)/\left<Q_0\right>\simeq 0.0711$, take place at $a\simeq 0.0$, $b\simeq 2.8$, which agrees well with the above mentioned best-fit values of those parameters. At the same time, the region of low $\sigma(Q_0)/\left<Q_0\right>$ values in the $(a, b)$ space does not fully coincide (although overlaps) with the regions of low $\eta^2$ metric (i.e., with the degeneracy stripes), and is apparently broader than the degeneracy stripes. In other words, the best-fit values of $a$ and $b$ providing the best model-to-observations agreement also imply that the corresponding best-fit $Q_0$ values are nearly the same for all emission frequencies, although the opposite is not always true. The apparent similarity between the obtained model fitting results at different frequencies confirms the validity of the adopted model as well as the accuracy of the SRH calibration. Capitalizing on these findings, we only use a subset of three distinct frequencies for other active regions and observation times.

\subsection{Dependence on the heating regime}\label{HeatingRegime}
Currently, the GX Simulator package \citep{Nita2023} contains the EBTEL lookup tables for the following heating regimes: 

a) ``Steady-state'' (i.e., constant with time) heating.

b) ``Impulsive'' heating by short periodical nanoflares with equal amplitudes occurring every $\tau=10^4$ s. 

c) Heating by stochastically occurring short nanoflares with the median time $\tau$ between them and the amplitudes following a power-law probability distribution defined by the slope $\alpha$. In turn, the median time $\tau$ between the nanoflares in a given magnetic loop is chosen to be proportional to the plasma cooling time (due to radiation and thermal conduction) $\tau_{\mathrm{c}}$ in that loop. Such a model is based on the assumption that for a more-or-less uniform sub-photospheric driver, the build-up time of free energy for a nanoflare is proportionally longer for longer loops, while the cooling time is also nearly proportional to the loop length \citep[cf.][]{Barnes2019, Mondal2024}. The GX Simulator package contains the EBTEL lookup tables for the stochastic nanoflare heating with the median times $\tau$ equal $0.2\tau_{\mathrm{c}}$ (high-frequency heating), $\tau_{\mathrm{c}}$ (intermediate-frequency heating), or $5\tau_{\mathrm{c}}$ (low-frequency heating), and the slopes $\alpha$ equal $-1$ or $-2.5$. Thus, there are eight supported heating regimes in total. The heating rate $Q$ used as an input for the EBTEL code is assumed to be time-averaged (over the timescales $\gg\tau$).

\begin{figure*}
\centerline{\large $\eta^2$ metric}\vspace{2pt}
\centerline{%
\includegraphics{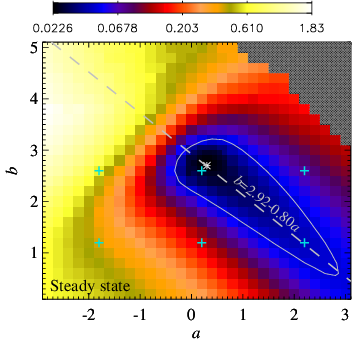}%
\includegraphics{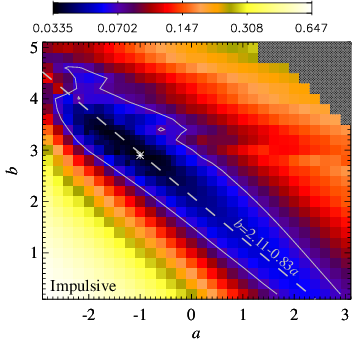}}
\centerline{%
\includegraphics{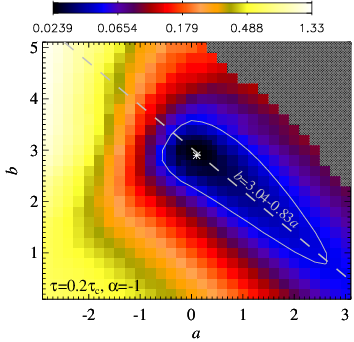}%
\includegraphics{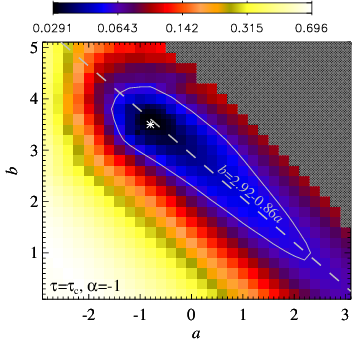}%
\includegraphics{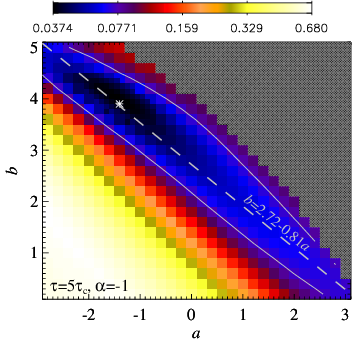}}
\centerline{%
\includegraphics{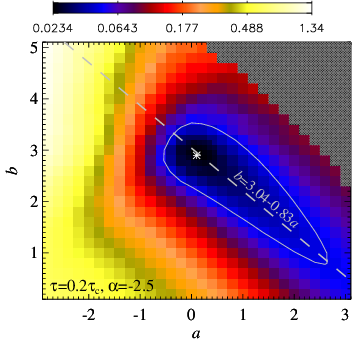}%
\includegraphics{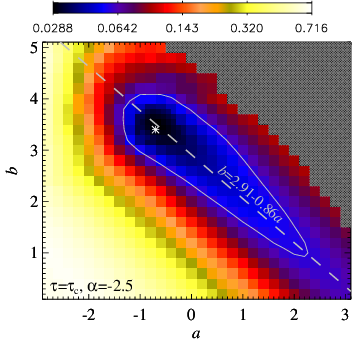}%
\includegraphics{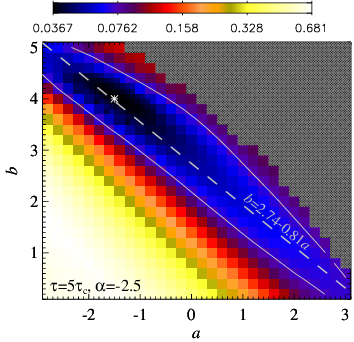}}\vspace{-7pt}
\caption{Maps of the fitting quality metric $\eta^2$ vs. the heating law parameters $(a, b)$, for AR 12396, at 2022-01-31 04:20 UT, at the frequency of 5.80 GHz, for different EBTEL heating regimes. The light-blue crosses in the upper-left panel mark some of the $(a, b)$ combinations chosen for a detailed model-to-observations comparison in Appendix \protect\ref{MultiImageComparison}. Other symbols, lines, and contours in the maps are the same as in Figure \protect\ref{FigMultiFreqs}.}
\label{FigMultiEBTEL}
\end{figure*}

\subsubsection{Qualitative comparison}
Figure \ref{FigMultiEBTEL} compares the $a{-}b$ diagrams, or maps of the fitting quality metric $\eta^2$ vs. the heating law parameters $a$ and $b$, for different EBTEL heating regimes. The diagrams were obtained for the same active region as in the previous examples (AR 12396 at 2022-01-31 04:20 UT), at the frequency of 5.80 GHz. One can see that the presented solutions of the forward-fitting problem have the same structure as described above, i.e., the regions of low $\eta^2$ metric form well-defined ``degeneracy stripes'' in the $(a, b)$ space. These degeneracy stripes have nearly the same locations and orientations (i.e., slopes $\delta$ and intercepts $b_0$) for most of the considered heating regimes. The impulsive heating regime stands slightly apart: the corresponding degeneracy stripe is parallel to those for other heating regimes, but is located a bit lower in the $a{-}b$ diagram (i.e., for the same values of $a$, the best-fit values of $b$ are lower by $\sim 0.8$); this behaviour was observed in other active regions as well.

At the same time, the profiles of the fitting quality metric along the degeneracy stripes are much more dependent on the heating regime: e.g., for the steady-state heating and heating by stochastic nanoflares with $\tau=0.2\tau_{\mathrm{c}}$, the stripes end abruptly at $a\simeq -0.5$ (when moving to upper-left), while for the heating by stochastic nanoflares with $\tau=\tau_{\mathrm{c}}$, impulsive heating, and, especially, heating by stochastic nanoflares with $\tau=5\tau_{\mathrm{c}}$, the stripes extend much farther into the domain of negative $a$ values. The best-fit values of the power-law indices $a$ and $b$ vary with the heating regime as well, from $(a, b)\simeq (0.3, 2.7)$ for the steady-state heating to $(a, b)\simeq (-1.5, 4.0)$ for the heating by stochastic nanoflares with $\tau=5\tau_{\mathrm{c}}$ and $\alpha=-2.5$. For the case presented in Figure \ref{FigMultiEBTEL}, the best agreement of the model with observations (with the lowest fitting quality metric of $\eta^2_{\min}\simeq 0.0226$) is achieved for the steady-state heating regime, while the low-frequency heating by stochastic nanoflares with $\tau=5\tau_{\mathrm{c}}$ and $\alpha=-1$ is the least favourable heating regime (with $\eta^2_{\min}\simeq 0.0374$). 

\begin{table*}
\caption{Characteristics of the best-fit coronal heating models for different active regions and different EBTEL heating regimes, at the selected frequencies: the best-fit power-law indices $a$ and $b$, and the corresponding values of the $\eta^2$ metric ($\eta^2_{\min}$, in percent). The minimum $\eta^2_{\min}$ values for each active region at each frequency are underlined.}
\label{TabMultiEBTEL}
\setlength{\tabcolsep}{1.1pt}
\begin{tabular}{cc}
\movetableright=-4.0cm
\renewcommand{\arraystretch}{0.97}
\begin{tabular}[t]{ccCCCcCCCcCCC}
\hline\hline
 & & \multicolumn{3}{c}{$\sim 6$ GHz} & & \multicolumn{3}{c}{$\sim 8$ GHz} & & \multicolumn{3}{c}{$\sim 10$ GHz}\\
\cline{3-5} \cline{7-9} \cline{11-13}
Model & & a & b & \eta^2_{\min} & & a & b & \eta^2_{\min} & & a & b & \eta^2_{\min}\\
\hline
 & & \multicolumn{11}{c}{AR12924, 2022-01-09 04:45}\\
Steady state & &  2.0 &  2.6 & \underline{ 1.80} & &  2.1 &  2.6 &  2.86 & &  1.5 &  2.7 & \underline{ 2.33}\\
Impulsive & &  0.2 &  3.5 &  2.39 & & -0.4 &  3.7 &  2.72 & & -0.3 &  3.9 &  2.52\\
$\tau{=}0.2\tau_{\mathrm{c}}$, $\alpha{=}{-}1$ & &  2.5 &  2.7 &  1.92 & &  1.7 &  2.8 &  2.68 & &  1.6 &  2.8 &  2.53\\
$\tau{=}0.2\tau_{\mathrm{c}}$, $\alpha{=}{-}2.5$ & &  2.7 &  2.6 &  1.91 & &  2.0 &  2.7 &  2.74 & &  1.5 &  2.8 &  2.45\\
$\tau{=}\tau_{\mathrm{c}}$, $\alpha{=}{-}1$ & &  1.5 &  3.6 &  2.53 & &  0.6 &  3.5 &  2.36 & &  0.9 &  3.5 &  2.83\\
$\tau{=}\tau_{\mathrm{c}}$, $\alpha{=}{-}2.5$ & &  0.7 &  3.3 &  2.48 & &  0.6 &  3.4 & \underline{ 2.30} & &  1.0 &  3.4 &  2.73\\
$\tau{=}5\tau_{\mathrm{c}}$, $\alpha{=}{-}1$ & & -0.0 &  3.9 &  3.95 & & -1.4 &  4.5 &  3.85 & &  0.2 &  4.7 &  3.79\\
$\tau{=}5\tau_{\mathrm{c}}$, $\alpha{=}{-}2.5$ & & -0.1 &  3.9 &  3.91 & & -1.5 &  4.5 &  3.68 & & -0.0 &  4.7 &  3.50\\
 & & \multicolumn{11}{c}{AR12936, 2022-01-31 04:20}\\
Steady state & &  0.3 &  2.7 & \underline{ 2.26} & & -0.1 &  2.5 &  2.84 & & -0.2 &  2.5 &  3.18\\
Impulsive & & -1.0 &  2.9 &  3.35 & & -1.5 &  2.9 &  3.47 & & -0.6 &  2.3 &  3.75\\
$\tau{=}0.2\tau_{\mathrm{c}}$, $\alpha{=}{-}1$ & &  0.1 &  2.9 &  2.39 & & -0.3 &  2.7 &  2.81 & & -0.4 &  2.7 &  3.18\\
$\tau{=}0.2\tau_{\mathrm{c}}$, $\alpha{=}{-}2.5$ & &  0.1 &  2.9 &  2.34 & & -0.3 &  2.7 & \underline{ 2.71} & & -0.4 &  2.7 &  3.13\\
$\tau{=}\tau_{\mathrm{c}}$, $\alpha{=}{-}1$ & & -0.8 &  3.5 &  2.91 & & -1.1 &  3.3 &  3.18 & & -1.1 &  3.4 &  3.07\\
$\tau{=}\tau_{\mathrm{c}}$, $\alpha{=}{-}2.5$ & & -0.7 &  3.4 &  2.88 & & -1.1 &  3.3 &  3.06 & & -1.0 &  3.3 & \underline{ 2.98}\\
$\tau{=}5\tau_{\mathrm{c}}$, $\alpha{=}{-}1$ & & -1.4 &  3.9 &  3.74 & & -1.2 &  3.5 &  4.23 & & -0.1 &  2.8 &  3.59\\
$\tau{=}5\tau_{\mathrm{c}}$, $\alpha{=}{-}2.5$ & & -1.5 &  4.0 &  3.67 & & -1.2 &  3.5 &  4.17 & & -0.3 &  2.8 &  3.58\\
 & & \multicolumn{11}{c}{AR13007, 2022-05-14 03:25}\\
Steady state & & -0.1 &  2.5 & \underline{ 2.89} & & -0.5 &  2.1 & \underline{ 2.70} & & -0.1 &  2.1 & \underline{ 6.18}\\
Impulsive & & -1.4 &  2.8 &  4.46 & & -2.0 &  2.8 &  3.90 & & -1.5 &  2.5 & 10.66\\
$\tau{=}0.2\tau_{\mathrm{c}}$, $\alpha{=}{-}1$ & & -0.3 &  2.7 &  3.05 & & -0.7 &  2.2 &  3.12 & & -0.2 &  2.2 &  7.45\\
$\tau{=}0.2\tau_{\mathrm{c}}$, $\alpha{=}{-}2.5$ & & -0.3 &  2.7 &  2.99 & & -0.6 &  2.2 &  3.03 & & -0.2 &  2.2 &  7.14\\
$\tau{=}\tau_{\mathrm{c}}$, $\alpha{=}{-}1$ & & -1.3 &  3.4 &  4.01 & & -1.1 &  2.7 &  4.59 & & -0.7 &  2.6 & 12.68\\
$\tau{=}\tau_{\mathrm{c}}$, $\alpha{=}{-}2.5$ & & -1.2 &  3.3 &  3.93 & & -1.1 &  2.7 &  4.48 & & -0.7 &  2.5 & 12.18\\
$\tau{=}5\tau_{\mathrm{c}}$, $\alpha{=}{-}1$ & & -1.8 &  3.7 &  5.41 & & -1.6 &  3.1 &  5.27 & & -1.8 &  3.5 & 13.65\\
$\tau{=}5\tau_{\mathrm{c}}$, $\alpha{=}{-}2.5$ & & -1.8 &  3.7 &  5.30 & & -1.7 &  3.2 &  5.20 & & -1.8 &  3.5 & 13.53\\
 & & \multicolumn{11}{c}{AR13234, 2023-02-26 03:40}\\
Steady state & & -0.0 &  2.5 & \underline{ 3.79} & & -0.2 &  2.6 &  6.96 & &  0.1 &  2.5 & \underline{11.37}\\
Impulsive & & -2.0 &  2.9 &  4.19 & & -2.7 &  3.4 & \underline{ 5.46} & & -2.0 &  3.3 & 13.77\\
$\tau{=}0.2\tau_{\mathrm{c}}$, $\alpha{=}{-}1$ & & -0.1 &  2.6 &  4.01 & & -0.3 &  2.7 &  7.08 & & -0.0 &  2.6 & 11.89\\
$\tau{=}0.2\tau_{\mathrm{c}}$, $\alpha{=}{-}2.5$ & & -0.1 &  2.6 &  3.98 & & -0.3 &  2.7 &  7.02 & & -0.0 &  2.6 & 11.75\\
$\tau{=}\tau_{\mathrm{c}}$, $\alpha{=}{-}1$ & & -1.2 &  3.1 &  4.42 & & -1.2 &  3.2 &  6.50 & & -0.6 &  3.0 & 13.68\\
$\tau{=}\tau_{\mathrm{c}}$, $\alpha{=}{-}2.5$ & & -0.8 &  2.9 &  4.48 & & -1.0 &  3.1 &  6.73 & & -0.4 &  2.9 & 13.68\\
$\tau{=}5\tau_{\mathrm{c}}$, $\alpha{=}{-}1$ & & -1.2 &  3.1 &  4.77 & & -3.0 &  4.2 &  5.73 & & -0.3 &  2.8 & 15.71\\
$\tau{=}5\tau_{\mathrm{c}}$, $\alpha{=}{-}2.5$ & & -1.2 &  3.1 &  4.75 & & -3.0 &  4.2 &  5.72 & & -0.8 &  3.1 & 15.61\\
\hline
\end{tabular} &
\movetableright=-3.0cm
\renewcommand{\arraystretch}{0.97}
\begin{tabular}[t]{ccCCCcCCCcCCC}
\hline\hline
 & & \multicolumn{3}{c}{$\sim 6$ GHz} & & \multicolumn{3}{c}{$\sim 8$ GHz} & & \multicolumn{3}{c}{$\sim 10$ GHz}\\
\cline{3-5} \cline{7-9} \cline{11-13}
Model & & a & b & \eta^2_{\min} & & a & b & \eta^2_{\min} & & a & b & \eta^2_{\min}\\
\hline
 & & \multicolumn{11}{c}{AR13245, 2023-03-09 04:00}\\
Steady state & & -0.2 &  2.3 & \underline{ 0.47} & & -0.6 &  2.5 & \underline{ 0.54} & & -0.4 &  2.6 &  1.35\\
Impulsive & & -0.9 &  2.1 &  0.86 & & -1.1 &  2.4 &  0.67 & & -2.3 &  3.3 & \underline{ 0.77}\\
$\tau{=}0.2\tau_{\mathrm{c}}$, $\alpha{=}{-}1$ & & -0.3 &  2.4 &  0.56 & & -0.7 &  2.6 &  0.56 & & -0.4 &  2.7 &  1.27\\
$\tau{=}0.2\tau_{\mathrm{c}}$, $\alpha{=}{-}2.5$ & & -0.3 &  2.4 &  0.54 & &  0.8 &  1.8 &  0.73 & & -0.3 &  2.6 &  1.29\\
$\tau{=}\tau_{\mathrm{c}}$, $\alpha{=}{-}1$ & & -0.6 &  2.6 &  0.84 & & -0.4 &  2.6 &  0.78 & & -1.7 &  3.6 &  0.85\\
$\tau{=}\tau_{\mathrm{c}}$, $\alpha{=}{-}2.5$ & & -0.3 &  2.4 &  0.82 & & -1.0 &  2.9 &  0.62 & & -1.1 &  3.2 &  0.98\\
$\tau{=}5\tau_{\mathrm{c}}$, $\alpha{=}{-}1$ & & -0.9 &  2.8 &  0.98 & & -1.2 &  3.1 &  0.83 & & -3.6 &  4.7 &  0.86\\
$\tau{=}5\tau_{\mathrm{c}}$, $\alpha{=}{-}2.5$ & & -0.9 &  2.8 &  0.97 & & -1.2 &  3.1 &  0.82 & & -3.6 &  4.7 &  0.84\\
 & & \multicolumn{11}{c}{AR13315, 2023-05-28 03:55}\\
Steady state & & -0.5 &  3.0 & \underline{ 3.62} & & -0.5 &  2.7 & \underline{ 7.27} & & -0.4 &  2.6 & \underline{ 9.29}\\
Impulsive & & -1.6 &  3.0 &  5.44 & & -0.4 &  2.0 &  8.52 & & -0.0 &  1.7 & 11.17\\
$\tau{=}0.2\tau_{\mathrm{c}}$, $\alpha{=}{-}1$ & & -0.8 &  3.2 &  3.86 & & -0.7 &  2.9 &  7.42 & & -0.6 &  2.7 &  9.65\\
$\tau{=}0.2\tau_{\mathrm{c}}$, $\alpha{=}{-}2.5$ & & -0.8 &  3.2 &  3.79 & & -0.6 &  2.8 &  7.38 & & -0.7 &  2.8 &  9.54\\
$\tau{=}\tau_{\mathrm{c}}$, $\alpha{=}{-}1$ & & -1.7 &  3.7 &  5.20 & & -0.6 &  2.7 &  8.43 & &  0.2 &  2.1 & 10.84\\
$\tau{=}\tau_{\mathrm{c}}$, $\alpha{=}{-}2.5$ & & -1.5 &  3.6 &  5.10 & & -0.4 &  2.6 &  8.35 & &  0.1 &  2.2 & 10.73\\
$\tau{=}5\tau_{\mathrm{c}}$, $\alpha{=}{-}1$ & & -1.9 &  3.8 &  6.78 & & -0.3 &  2.5 &  9.90 & & -0.1 &  2.4 & 11.67\\
$\tau{=}5\tau_{\mathrm{c}}$, $\alpha{=}{-}2.5$ & & -1.9 &  3.8 &  6.66 & & -0.6 &  2.7 &  9.78 & & -0.3 &  2.5 & 11.60\\
 & & \multicolumn{11}{c}{AR13372, 2023-07-17 03:45}\\
Steady state & & -0.7 &  2.8 & \underline{ 3.70} & &  1.6 &  1.9 & \underline{ 5.84} & &  2.2 &  1.6 & \underline{ 5.70}\\
Impulsive & & -1.5 &  2.6 &  4.78 & &  2.4 &  0.9 &  7.07 & &  2.1 &  1.1 &  6.08\\
$\tau{=}0.2\tau_{\mathrm{c}}$, $\alpha{=}{-}1$ & & -0.9 &  3.0 &  3.82 & &  1.7 &  1.9 &  6.08 & &  1.2 &  2.1 &  5.77\\
$\tau{=}0.2\tau_{\mathrm{c}}$, $\alpha{=}{-}2.5$ & & -0.8 &  2.9 &  3.81 & &  1.7 &  1.9 &  6.02 & &  1.2 &  2.1 &  5.75\\
$\tau{=}\tau_{\mathrm{c}}$, $\alpha{=}{-}1$ & & -1.9 &  3.5 &  4.76 & &  1.8 &  1.8 &  7.13 & &  1.3 &  2.1 &  6.35\\
$\tau{=}\tau_{\mathrm{c}}$, $\alpha{=}{-}2.5$ & & -1.7 &  3.4 &  4.72 & &  2.1 &  1.7 &  7.02 & &  1.3 &  2.1 &  6.23\\
$\tau{=}5\tau_{\mathrm{c}}$, $\alpha{=}{-}1$ & & -1.7 &  3.4 &  5.55 & &  2.2 &  1.6 &  8.51 & &  1.5 &  2.0 &  7.85\\
$\tau{=}5\tau_{\mathrm{c}}$, $\alpha{=}{-}2.5$ & & -1.9 &  3.5 &  5.51 & &  2.2 &  1.6 &  8.42 & &  1.5 &  2.0 &  7.72\\
 & & \multicolumn{11}{c}{AR13559, 2024-01-24 04:15}\\
Steady state & & -0.0 &  2.5 & \underline{ 4.01} & &  0.2 &  2.5 & \underline{ 3.68} & &  0.2 &  2.4 & \underline{ 6.21}\\
Impulsive & & -1.1 &  2.5 &  6.01 & & -0.4 &  2.2 &  5.23 & &  0.6 &  1.6 &  7.94\\
$\tau{=}0.2\tau_{\mathrm{c}}$, $\alpha{=}{-}1$ & & -0.1 &  2.6 &  4.52 & & -0.0 &  2.6 &  4.01 & &  0.2 &  2.4 &  6.71\\
$\tau{=}0.2\tau_{\mathrm{c}}$, $\alpha{=}{-}2.5$ & & -0.1 &  2.6 &  4.41 & & -0.0 &  2.6 &  3.96 & &  0.2 &  2.4 &  6.62\\
$\tau{=}\tau_{\mathrm{c}}$, $\alpha{=}{-}1$ & & -0.8 &  2.9 &  6.02 & & -0.2 &  2.7 &  4.99 & &  0.2 &  2.4 &  8.12\\
$\tau{=}\tau_{\mathrm{c}}$, $\alpha{=}{-}2.5$ & & -0.6 &  2.8 &  5.93 & & -0.2 &  2.7 &  4.88 & &  0.2 &  2.4 &  7.99\\
$\tau{=}5\tau_{\mathrm{c}}$, $\alpha{=}{-}1$ & & -1.1 &  3.1 &  6.65 & & -0.6 &  2.9 &  5.66 & &  1.0 &  1.9 &  8.62\\
$\tau{=}5\tau_{\mathrm{c}}$, $\alpha{=}{-}2.5$ & & -1.1 &  3.1 &  6.61 & & -0.4 &  2.8 &  5.64 & &  1.0 &  1.9 &  8.59\\
\hline
\end{tabular}
\end{tabular}
\end{table*}

\subsubsection{Statistical analysis}
We performed a similar analysis for all eight considered active regions (at one representative time for each active region), at three selected frequencies (5.80, 7.80, and 9.80 GHz for the active regions observed in years $2022-2023$, and 6.00, 8.00, and 10.00 GHz for the active region observed in year 2024). The results, i.e., the best-fit values of $a$ and $b$ indices and the corresponding (minimum) values of the fitting quality metric $\eta^2_{\min}$, are summarized in Table \ref{TabMultiEBTEL}. It follows from the Table that:

a) The steady-state heating regime looks the most favourable one, i.e., it provides the best agreement between the model and observations (in terms of the $\eta^2$ metric) for the majority of active regions and frequencies. This occurred in all eight considered active regions at the frequency of $\sim 6$ GHz, in five active regions at $\sim 8$ GHz, and in six active regions at $\sim 10$ GHz, i.e., in 19 cases out of 24. In four active regions (AR 13007, 13315, 13372, and 13559) out of eight, the steady-state heating regime was the most favourable one at all three considered frequencies.

b) High-frequency heating by stochastic nanoflares (with the median cadence of $\tau=0.2\tau_{\mathrm{c}}$) typically provided the results very similar to those for the steady-state heating, with the model-to-observations fitting quality being only slightly worse, and sometimes even better. In one case (in AR 12936, at the frequency of 7.80 GHz), heating by stochastic nanoflares with $\tau=0.2\tau_{\mathrm{c}}$ and $\alpha=-2.5$ was found to be the most favourable heating regime.

c) The next group of heating regimes that provided mutually similar model-to-observations fitting quality comprises the intermediate-frequency stochastic heating (with the median cadence of $\tau=\tau_{\mathrm{c}}$) and the impulsive heating by periodic nanoflares with the cadence of $10^4$ s. Typically, these heating regimes were less successful in reproducing the observed microwave images than the steady-state or high-frequency stochastic heating, with four exceptions: the heating by stochastic nanoflares with $\tau=\tau_{\mathrm{c}}$ and $\alpha=-2.5$ was the most favourable heating regime in AR 12924 at the frequency of 7.80 GHz and in AR 12936 at 9.80 GHz, while the impulsive heating was the most favourable in AR 13234 at the frequency of 7.80 GHz and in AR 13245 at 9.80 GHz. We note that at other frequencies in the mentioned four active regions, either steady-state or high-frequency stochastic heating regimes were the most favourable ones.

d) Low-frequency heating by stochastic nanoflares (with the median cadence of $\tau=5\tau_{\mathrm{c}}$) never was the most favourable heating regime. Typically, it provided the worst model-to-observations agreement among all considered heating regimes, with two exceptions: in AR 13234 at the frequency of 7.80 GHz and in AR 13245 at 9.80 GHz, where the impulsive heating regime was the most favourable one, heating by stochastic nanoflares with $\tau=5\tau_{\mathrm{c}}$ and $\alpha=-2.5$ was the second best.

e) If we compare the regimes of heating by stochastic nanoflares with the same median cadence $\tau$ but different power-law slopes $\alpha$, the regimes with a steeper slope of $\alpha=-2.5$ typically provided a slightly better agreement of the model with observations (in 65 pairs out of 72). Hence the stochastic heating regimes dominated by weaker nanoflares look more favourable.

f) The particular best-fit parameters $a$ and $b$ of the coronal heating model were dependent on the chosen heating regime. Typically, the steady-state and high-frequency stochastic heating regimes provided the largest $a$ and smallest $b$ values, while the low-frequency stochastic heating regimes provided the smallest $a$ and largest $b$ values (i.e., in the same way as in the case shown in Figure \ref{FigMultiEBTEL}).

\subsubsection{The most likely heating regime(s)}
Summarizing the above results and assuming that the best model-to-observations agreement corresponds to the most relevant heating regime, we conclude that in the considered active regions the coronal plasma was heated predominantly either due to a steady-state (continuous) energy release process, or by frequently occurring weak nanoflares with a typical cadence of $\tau\ll\tau_{\mathrm{c}}$. Actually, the steady-state and high-frequency nanoflare heating models proved to be almost indistinguishable from each other. At the same time, in a few active regions at certain emission frequencies, the impulsive or intermediate-frequency heating models with the nanoflare cadences of $\tau=10^4$ s or $\tau=\tau_{\mathrm{c}}$, respectively, provided a better agreement with the observations. These apparent changes of the most favourable heating model with frequency (i.e., at the corresponding gyrolayers) might be due to disproportionally large contribution of specific selectively heated coronal loops \citep[cf.][]{Fleishman2025}, where the heating regime could be different from the diffuse component of the active region. The low-frequency nanoflare heating models with a typical cadence of $\tau\gg\tau_{\mathrm{c}}$ were never the best in our set of the data.

\subsection{Diversity in the coronal heating model parameters}
We now analyze in more detail how the best-fit coronal heating model parameters varied from one active region to another, and also with time. We restrict the analysis to the steady-state heating regime because, as demonstrated above, it typically provided the best (or one of the best) agreement between the model and observations; in addition, the structure of the solutions (i.e., locations and orientations of the ``degeneracy stripes'') has been found to be nearly the same for all heating regimes.

\begin{table*}
\caption{Best-fit parameters of the coronal heating model for different active regions at different times, for the steady-state EBTEL heating regime, at the selected frequencies: the best-fit power-law indices ($a$ and $b$), the corresponding total heating rates ($P_{\mathrm{AR}}$, in units of $10^{25}$ erg $\textrm{s}^{-1}$), and the slopes ($\delta$), intercepts ($b_0$), and widths ($w$) of the degeneracy stripes in the $a{-}b$ diagrams.}
\label{TabMultiARs}
\setlength{\tabcolsep}{1.7pt}
\movetableright=-3.7cm
\renewcommand{\arraystretch}{0.97}
\begin{tabular}{ccCCCCCCcCCCCCCcCCCCCC}
\hline\hline
 & & \multicolumn{6}{c}{$\sim 6$ GHz} & & \multicolumn{6}{c}{$\sim 8$ GHz} & & \multicolumn{6}{c}{$\sim 10$ GHz}\\
\cline{3-8} \cline{10-15} \cline{17-22}
Time, UT & & a & b & P_{\mathrm{AR}} & \delta & b_0 & w & & a & b & P_{\mathrm{AR}} & \delta & b_0 & w & & a & b & P_{\mathrm{AR}} & \delta & b_0 & w\\
\hline
2022-01-09 04:45 & &  2.0 &  2.6 &  0.88 & -0.16 & 2.94 & 0.95 & &  2.1 &  2.6 &  0.67 & -0.27 & 3.15 & 1.27 & &  1.5 &  2.7 &  0.59 & -0.19 & 2.96 & 0.97\\
2022-01-09 05:35 & &  2.2 &  2.5 &  0.74 & -0.27 & 3.09 & 1.42 & &  1.4 &  2.5 &  0.95 & -0.24 & 2.83 & 3.90 & &  0.9 &  2.6 &  0.89 & -0.49 & 3.05 & 1.86\\
2022-01-10 05:25 & &  1.2 &  2.3 &  1.53 & -0.47 & 2.93 & 3.88 & &  0.2 &  2.0 &  1.64 & -0.93 & 2.05 & 6.25 & &  0.7 &  0.5 &  1.46 & -1.07 & 1.18 & 8.71\\
2022-01-10 06:10 & &  1.5 &  2.3 &  1.43 & -0.43 & 2.99 & 3.89 & &  0.2 &  2.2 &  1.50 & -0.78 & 2.41 & 4.33 & &  0.3 &  2.0 &  1.35 & -0.73 & 2.19 & 4.48\\
2022-01-30 04:15 & & -0.4 &  1.8 & 14.53 & -1.14 & 1.46 & 1.18 & & -0.3 &  1.7 & 12.25 & -1.15 & 1.40 & 0.95 & & -0.1 &  1.6 & 12.08 & -1.22 & 1.46 & 0.91\\
2022-01-30 06:25 & & -0.8 &  1.6 & 21.19 & -1.01 & 0.79 & 1.43 & & -1.4 &  1.9 & 24.93 & -1.15 & 0.32 & 1.60 & & -1.4 &  2.1 & 24.16 & -1.17 & 0.54 & 1.32\\
2022-01-31 04:20 & &  0.3 &  2.7 &  1.39 & -0.80 & 2.92 & 1.19 & & -0.1 &  2.5 &  2.20 & -0.87 & 2.47 & 1.13 & & -0.2 &  2.5 &  2.43 & -0.92 & 2.38 & 1.28\\
2022-01-31 06:15 & &  0.4 &  2.6 &  1.74 & -0.89 & 2.98 & 2.56 & & -0.2 &  2.6 &  2.47 & -0.88 & 2.54 & 1.68 & & -0.1 &  2.6 &  2.27 & -0.84 & 2.53 & 1.13\\
2022-05-14 03:25 & & -0.1 &  2.5 &  3.56 & -0.88 & 2.59 & 1.81 & & -0.5 &  2.1 & 10.54 & -0.90 & 1.76 & 1.30 & & -0.1 &  2.1 &  5.06 & -0.95 & 1.95 & 1.45\\
2022-05-14 06:35 & &  0.3 &  2.1 &  2.96 & -0.74 & 2.35 & 2.90 & & {>} 7.6 & {<}-4.0 & \nodata & -0.87 & 2.56 & 2.84 & & {>} 8.9 & {<}-5.6 & \nodata & -0.92 & 2.50 & 2.41\\
2022-05-15 03:45 & &  0.3 &  2.2 &  3.10 & -0.51 & 2.34 & 0.82 & &  0.1 &  2.3 &  3.26 & -0.47 & 2.37 & 1.12 & &  0.2 &  2.3 &  2.57 & -0.47 & 2.33 & 0.86\\
2022-05-15 06:20 & &  1.0 &  1.8 &  2.80 & -0.58 & 2.40 & 0.92 & & {>} 6.9 & {<}-1.8 & \nodata & -0.57 & 2.25 & 1.07 & &  4.7 & -0.5 &  0.72 & -0.59 & 2.22 & 1.61\\
2023-02-26 03:40 & &  0.0 &  2.5 &  2.30 & -0.56 & 2.52 & 0.76 & & -0.2 &  2.6 &  2.19 & -0.55 & 2.49 & 0.68 & &  0.1 &  2.5 &  1.67 & -0.56 & 2.59 & 0.89\\
2023-02-26 06:40 & & -0.3 &  2.4 &  4.73 & -0.59 & 2.27 & 0.53 & & -0.2 &  2.5 &  3.53 & -0.55 & 2.42 & 0.39 & &  0.2 &  2.4 &  2.39 & -0.54 & 2.54 & 0.45\\
2023-02-27 03:45 & & -0.5 &  2.6 &  2.73 & -0.58 & 2.35 & 0.62 & & -0.6 &  2.7 &  2.38 & -0.59 & 2.40 & 0.71 & & -0.1 &  2.5 &  1.59 & -0.61 & 2.44 & 0.91\\
2023-02-27 06:45 & & -0.2 &  2.6 &  3.37 & -0.63 & 2.51 & 0.86 & & -0.6 &  2.7 &  2.85 & -0.62 & 2.37 & 0.72 & & -0.4 &  2.7 &  1.86 & -0.63 & 2.49 & 0.86\\
2023-03-09 04:00 & & -0.2 &  2.3 &  2.87 & -0.54 & 2.21 & 0.30 & & -0.6 &  2.5 &  3.71 & -0.56 & 2.21 & 0.25 & & -0.4 &  2.6 &  2.60 & -0.54 & 2.32 & 1.41\\
2023-03-09 06:25 & & -0.7 &  2.4 &  6.58 & -0.54 & 2.00 & 0.35 & &  0.4 &  1.9 &  2.88 & -0.55 & 2.13 & 0.27 & &  4.2 &  0.1 &  1.05 & -0.56 & 2.38 & 0.51\\
2023-03-10 03:45 & &  0.4 &  1.8 &  5.98 & -0.52 & 2.01 & 0.40 & & -0.5 &  2.3 &  7.36 & -0.53 & 2.04 & 0.60 & & -0.2 &  2.3 &  5.15 & -0.49 & 2.09 & 2.07\\
2023-03-10 06:20 & &  1.7 &  1.2 &  5.11 & -0.50 & 2.02 & 0.45 & & -0.3 &  2.2 &  7.69 & -0.52 & 2.02 & 0.48 & & -0.5 &  2.6 &  6.41 & \nodata & \nodata & \nodata\\
2023-05-27 03:50 & &  2.8 &  0.5 &  1.02 & -0.84 & 2.85 & 1.73 & & {>} 5.7 & {<}-2.3 & \nodata & -0.88 & 2.68 & 1.77 & &  4.4 & -1.4 &  2.02 & -0.88 & 2.43 & 1.56\\
2023-05-27 06:30 & &  1.7 &  1.7 &  0.69 & -0.86 & 3.10 & 1.21 & &  4.4 & -1.0 &  0.87 & -0.90 & 2.92 & 1.41 & &  4.8 & -1.5 &  1.13 & -0.94 & 2.98 & 1.59\\
2023-05-28 03:55 & & -0.5 &  3.0 &  2.76 & -0.69 & 2.74 & 0.96 & & -0.5 &  2.7 &  3.24 & -0.70 & 2.36 & 1.43 & & -0.4 &  2.6 &  3.39 & -0.70 & 2.31 & 1.49\\
2023-05-28 06:50 & & -0.1 &  2.7 &  2.54 & -0.66 & 2.66 & 1.22 & & -0.2 &  2.5 &  3.32 & -0.72 & 2.44 & 1.56 & & -0.1 &  2.4 &  3.44 & -0.73 & 2.36 & 1.22\\
2023-07-17 03:45 & & -0.7 &  2.8 &  3.14 & -0.46 & 2.50 & 0.81 & &  1.6 &  1.9 &  1.13 & -0.47 & 2.62 & 0.82 & &  2.2 &  1.6 &  0.70 & -0.47 & 2.61 & 2.72\\
2023-07-17 06:15 & & -0.7 &  2.7 &  3.73 & -0.51 & 2.44 & 0.90 & & -0.8 &  2.9 &  2.85 & -0.46 & 2.46 & 1.22 & &  1.3 &  1.9 &  0.99 & -0.49 & 2.50 & 4.33\\
2023-07-18 04:00 & &  0.4 &  2.5 &  3.28 & -0.53 & 2.77 & 1.13 & & -0.5 &  2.5 &  4.70 & -0.49 & 2.18 & 2.14 & & -0.9 &  2.5 &  4.65 & -0.44 & 1.92 & 6.99\\
2023-07-18 06:40 & &  0.3 &  2.7 &  3.35 & -0.53 & 2.93 & 1.40 & &  0.0 &  2.7 &  3.42 & -0.53 & 2.62 & 2.18 & & -0.6 &  2.7 &  2.94 & -0.50 & 2.43 & 6.98\\
2024-01-23 04:20 & & -0.1 &  2.6 &  4.79 & -0.70 & 2.52 & 1.54 & & -0.2 &  2.7 &  4.67 & -0.66 & 2.48 & 1.22 & & -0.1 &  2.5 &  5.61 & -0.66 & 2.40 & 1.53\\
2024-01-23 05:50 & & -0.1 &  2.4 &  5.28 & -0.72 & 2.34 & 1.83 & & -0.2 &  2.5 &  5.72 & -0.71 & 2.39 & 1.19 & &  3.5 & -0.2 &  2.71 & -0.71 & 2.32 & 1.26\\
2024-01-24 04:15 & &  0.0 &  2.5 &  3.49 & -0.68 & 2.50 & 1.06 & &  0.2 &  2.5 &  2.33 & -0.64 & 2.56 & 0.49 & &  0.2 &  2.4 &  2.32 & -0.65 & 2.50 & 0.55\\
2024-01-24 06:15 & & -0.2 &  2.7 &  3.46 & -0.64 & 2.54 & 0.88 & & -0.2 &  2.7 &  2.97 & -0.66 & 2.59 & 0.66 & & -0.2 &  2.7 &  2.55 & -0.67 & 2.55 & 0.61\\
\hline
Median & &  0.0 &  2.5 &  3.12 & -0.59 & 2.51 & 1.09 & & -0.2 &  2.5 &  2.97 & -0.63 & 2.41 & 1.20 & & -0.1 &  2.4 &  2.39 & -0.65 & 2.40 & 1.41\\
\hline
\end{tabular}
\end{table*}

Table \ref{TabMultiARs} summarizes the results of fitting the coronal heating model to the SRH observations for all considered active regions at all considered times (i.e., for 32 cases in total), at three selected frequencies; the corresponding $a{-}b$ diagrams are presented in Appendix D in the ApJ version of the paper. The Table presents the best-fit values of $a$ and $b$ indices and the corresponding total heating rates $P_{\mathrm{AR}}$ (see below), and the estimated characteristics of the degeneracy stripes in the $a{-}b$ diagrams: slopes $\delta$, intercepts $b_0$, and widths $w$ (at the $\eta^2=2\eta_{\min}^2$ level, see Appendix \ref{StripeFitting}). In four cases, the best-fit heating model parameters $(Q_0, a, b)$ were not found, because the required heating intensities $Q_0$ were located beyond the limits determined by the used EBTEL lookup table. Also, in one case the degeneracy stripe was poorly defined (see Section \ref{Outliers}), so that the general algorithm presented in Appendix \ref{StripeFitting} failed to find a reasonable straight-line fit to it.

\subsubsection{Typical characteristics and trends of the solutions}
Analysis of all selected observations has revealed that in most cases, the solutions of the forward-fitting problem demonstrated well-defined ``degeneracy stripes''. We define the stripe width $w$ as the maximum extent of the area with $\eta^2<2\eta^2_{\min}$ in the direction perpendicular to the stripe axis (see Appendix \ref{StripeFitting}). In the considered active regions, the degeneracy stripes had a typical (median) width of $w\simeq 1$, with $w<2$ in 81\% of cases and $w<3$ in 89\% of cases.

Typically, the degeneracy stripes at different frequencies had nearly the same parameters. Exceptions could be noticed in only a few cases, when the degeneracy stripes were too wide (like, e.g., in AR 12924 at 2022-01-10 05:25 UT), so that their parameters might be not reliably determined; such cases are discussed in Section \ref{Outliers}. 

Based on the visual analysis, we define the ``typical'' cases as those with the stripe widths of $w<3$. If we consider only those typical cases, the slopes $\delta$ of the degeneracy stripes decreased (steepened) monotonously with frequency in 11 cases, increased (flattened) monotonously with frequency in five cases, and demonstrated a mixed frequency trend in nine cases out of 25, i.e., the steepening trend prevailed slightly but was far from dominant; the variations $\max(\delta)-\min(\delta)$ did not exceed 0.18 and were typically much lower. For the same narrow degeneracy stripes, the intercepts $b_0$ increased monotonously with frequency in five cases, decreased monotonously with frequency in seven cases, and demonstrated a mixed frequency trend in 13 cases out of 25, i.e., there was no dominant frequency trend as well. These findings confirm the above conclusion that the coronal heating processes have nearly the same characteristics, and the relationship between the magnetic flux tube parameters is described by nearly the same scaling law within the considered range of the microwave emission frequencies (i.e., within the corresponding range of heights); the detected frequency dependences were not significant enough to make any physical conclusions.

\begin{figure}
\centerline{\includegraphics{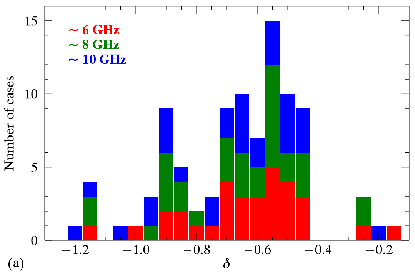}}
\centerline{\includegraphics{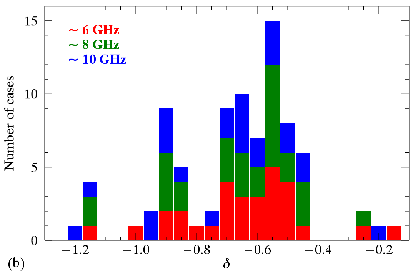}}
\caption{Histograms of the slopes $\delta$ of the best-fit (``degeneracy'') stripes in the $a{-}b$ diagrams, with the bin size of 0.05. The panels show the histograms for all considered active regions and times (a), and for those with relatively narrow degeneracy stripes ($w<3$) only (b).}
\label{FigChistograms}
\end{figure}

\subsubsection{Characteristics of the degeneracy stripes}\label{delta_b0_scalings}
The slopes $\delta$ of the degeneracy stripes had comparable values within any given active region at different times, but varied considerably from one active region to another. We remind that this slope characterizes the power-law relationship between the magnetic flux tube length and the average magnetic field in that tube, according to Equation (\ref{Bscale}). The obtained statistical distribution of the slopes/indices $\delta$ is shown in Figure \ref{FigChistograms}. One can see that the most probable value was $\delta\simeq -0.55$; other likely values were $\delta\simeq -(0.65-0.70)$ and $\delta\simeq -0.90$. Overall, $\delta$ varied from about $-1.2$ (in AR 12936 on 2022-01-30) to about $-0.2$ (in AR 12924 on 2022-01-09). The presence of several local maxima in the distribution of $\delta$ was likely caused by a limited sample of active regions in the presented study, while for a larger sample we can expect a more continuous asymmetric distribution looking like an envelope of the distribution shown in Figure \ref{FigChistograms}, with a peak near $\delta\simeq -0.5$. 

\begin{figure}
\centerline{\includegraphics{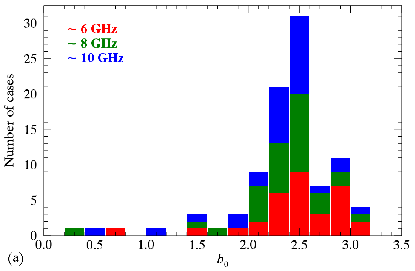}}
\centerline{\includegraphics{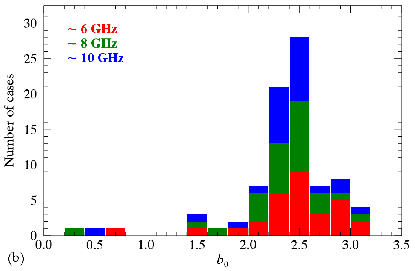}}
\caption{Histograms of the intercepts $b_0$ of the best-fit (``degeneracy'') stripes in the $a{-}b$ diagrams, with the bin size of 0.2. The panels show the histograms for all considered active regions and times (a), and for those with relatively narrow degeneracy stripes ($w<3$) only (b).}
\label{FigB0histograms}
\end{figure}

Figure \ref{FigB0histograms} demonstrates the statistical distribution of the intercepts $b_0$ of the degeneracy stripes. We remind that this parameter characterizes the effective dependence of the heating rate on the magnetic flux tube length, according to Equation (\ref{HeatingRate2}). One can see from the figure that the most probable value was $b_0\simeq 2.5$, and the majority of values were confined within the range from 1.6 to 3.2, with an approximately normal distribution. The only exception was AR 12396 on 2022-01-30, when the $b_0$ values were much lower: below 1.6 and down to 0.32. It is interesting to note that the same active region on the same date was also characterized by extremely steep slopes $\delta$ (down to $-1.2$).

\subsubsection{The best-fit $a$ and $b$ indices}
The most typical (median) best-fit values of the power-law indices $a$ and $b$ were $a\simeq -0.1$ and $b\simeq 2.5$. The fact that the most typical values of the $b$ and $b_0$ indices proved to be the same is not a coincidence: when the $a$ index is close to zero ($|a|\ll 1$, like in the mentioned most typical case), the original coronal heating model (\ref{HeatingRate}) is effectively reduced to  model (\ref{HeatingRate2}) involving the dependence on the magnetic flux tube length only, so that $b_0$ approaches $b$. At the same time, the best-fit indices $a$ and $b$ varied in a wide range across Table \ref{TabMultiARs}: the best-fit $a$ values varied from $-1.4$ to 8.9, and the best-fit $b$ values varied from 3.0 to $-5.6$, with the upper boundary of $a$ and lower boundary of $b$ formed due to the model restrictions (i.e., due to the limited EBTEL lookup table). Sometimes (e.g., in AR 13315 on 2023-05-27), the best-fit $(a, b)$ values varied considerably with frequency as well, ``jumping'' back and forth along the degeneracy stripe. 

\begin{figure}
\centerline{\includegraphics{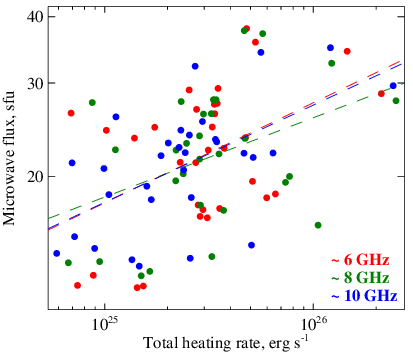}}
\caption{The microwave fluxes $I$ from the considered active regions (see Table \protect\ref{TabARlist}) vs. the estimated total heating rates $P_{\mathrm{AR}}$ (see Table \protect\ref{TabMultiARs}), at three selected frequencies. Dashed lines represent the power-law fits, with the correlation coefficients between $\log I$ and $\log P_{\mathrm{AR}}$ being 0.46, 0.40, and 0.58 at the frequencies of about 6, 8, and 10 GHz, respectively.}
\label{FigPIcorrelations}
\end{figure}

\subsubsection{Total heating rates}
The total heating rate within the active region $P_{\mathrm{AR}}$ (measured in erg $\textrm{s}^{-1}$) can be defined as
\begin{equation}\label{TotalHeatingRate}
P_{\mathrm{AR}}=\int Q\,\mathrm{d}V,
\end{equation}
where, for consistency with the microwave simulations, the integration should be performed over the volume elements associated with closed magnetic field lines and with the corresponding $Q$ and $L$ parameters within the EBTEL table. In our study, the total heating rates $P_{\mathrm{AR}}$ varied in a wide range from $5.88\times 10^{24}$ to $2.49\times 10^{26}$ erg $\textrm{s}^{-1}$, with the median value of $2.85\times 10^{25}$ erg $\textrm{s}^{-1}$. For each active region at each time, the total heating rates inferred from observations at different frequencies were mostly consistent with each other, without a definite frequency trend, thus confirming that the basic parameters of the coronal heating processes do not vary significantly within an active region. The total heating rates demonstrated a positive correlation with the microwave fluxes from the active regions (see Figure \ref{FigPIcorrelations}), i.e., higher fluxes typically required a stronger heating, although the scatter of values was significant. 

\subsubsection{Variations with time}\label{TimeVariations}
For each of the selected active regions, the observations have been analyzed at four different times, which allows us to examine how the magnetic field structure and the coronal heating processes varied with time. One can see from Table \ref{TabMultiARs} that for the observations performed at two different times on the same day, the inferred parameters of the coronal heating model were typically very similar, which indicates that both the magnetic field structures in the active regions (characterized by the index $\delta$) and the parameters of the plasma heating mechanism (characterized by the best-fit indices $a$ and $b$ and the total heating rate $P_{\mathrm{AR}}$) did not vary significantly on the timescales of a few hours. On the other hand, the observations performed on adjacent days revealed that the variations of the mentioned characteristics on the timescales of about one day could be more significant: e.g., the total heating rate $P_{\mathrm{AR}}$ sometimes varied by a factor of $\gtrsim 2$.

The most significant variations have been detected in AR 12396, that on 2022-01-30 demonstrated the total heating rate $\sim 10$ times higher than on the next day, 2022-01-31; there were strong variations in the slopes $\delta$ and, especially, intercepts $b_0$ of the degeneracy stripes in the $a{-}b$ diagrams as well. The \textit{SDO}/AIA EUV observations, too, revealed a noticeable change of morphology in AR 12936 between 2022-01-30 and 2022-01-31 (see the corresponding figures in the figure set in Appendix D in the ApJ version of the paper), which might reflect the processes of dissipation of non-potential magnetic energy. During the considered time interval, AR 12936 produced a number of flares, including eight flares of approximately C1 GOES class and several flares undetected by GOES (according to the Heliophysics Events Knowledgebase). However, in this respect AR 12936 was not qualitatively different from other analyzed active regions: e.g., AR 13245 and AR 13359 each produced the same number (eight) of even stronger flares (up to M1 class) between the observations on adjacent days, but demonstrated much weaker variations of the inferred parameters of the coronal heating model.

\begin{figure*}
\centerline{\large $\eta^2$ metric}\vspace{3pt}
\centerline{
\includegraphics{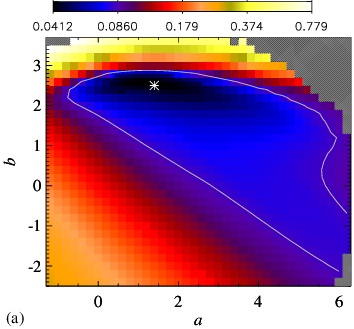}
\includegraphics{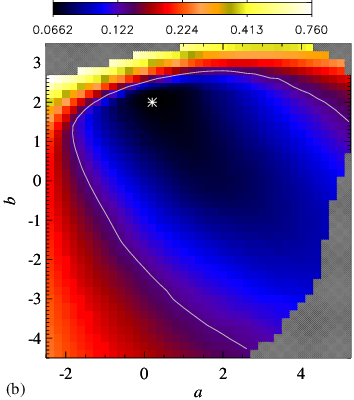}}
\centerline{
\includegraphics{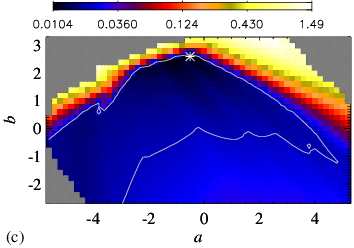}
\includegraphics{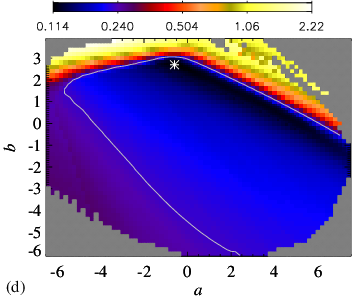}}
\caption{Examples of atypical maps of the fitting quality metric $\eta^2$ vs. the heating law parameters $(a, b)$, for the following active regions, times, and frequencies: (a) AR 12924, 2022-01-09 05:35 UT, 7.80 GHz; (b) AR 12924, 2023-01-10 05:25 UT, 7.80 GHz; (c) AR 13245, 2023-03-10 06:20 UT, 9.80 GHz; (d) AR 13372, 2023-07-18 06:40 UT, 9.80 GHz. All maps correspond to the steady-state EBTEL heating regime. The minima of $\eta^2$ are marked by asterisks, and solid white lines represent the contours of $\eta^2=2\eta^2_{\min}$.}
\label{FigOutliers}
\end{figure*}

\subsubsection{The ``atypical'' cases}\label{Outliers}
As has been said above, in most of the analyzed cases the solutions of the forward-fitting problem had a similar structure, when the model parameters providing a good model-to-observations agreement formed narrow straight stripes in the $(a, b)$ space. However, in a few cases the solutions deviated considerably from the mentioned typical structure, i.e., the ``degeneracy stripes'' could not be clearly identified or were too wide, which made determining the corresponding scaling-law indices $\delta$ and $b_0$ more difficult or impossible. A few examples of such atypical $a{-}b$ diagrams are shown in Figure \ref{FigOutliers}. 

In particular, in the diagram in Figure \ref{FigOutliers}a, the region of small values of the fitting quality metric $\eta^2$ has a tadpole-like shape, with the width in $b$-direction firstly increasing and then sharply decreasing with an increase of $a$; one can notice indications of not one but two degeneracy stripes with different slopes (the lower stripe is longer) intersecting at $(a, b)\simeq (1.0, 2.5)$. Such a structure in the diagram can reflect the presence of two populations of magnetic flux tubes described by  scaling law (\ref{Bscale}) with different $\delta$ indices. In the diagram in Figure \ref{FigOutliers}b, the region of small $\eta^2$ values forms a very wide and noticeably curved stripe that might be formed due to overlapping of multiple (more than two) narrower stripes, or if the parameters of the magnetic flux tubes followed a more-or-less continuous distribution with the equivalent values of the $\delta$ index in Equation (\ref{Bscale}) varying in a broad range (which actually means that  scaling law (\ref{Bscale}) was not an adequate model). In the diagrams in Figures \ref{FigOutliers}c-d, the stripes of small $\eta^2$ values are wide and asymmetric, with sharp upper edges (when $b$ increases) and a slow gradual increase of $\eta^2$ when $b$ decreases; such structures might be formed due to overlapping of multiple narrow degeneracy stripes that, if extended to infinity, would intersect at a point beyond the upper/left boundaries of the displayed maps (with small $a$ and large $b$ values).

We conclude that the atypical $a{-}b$ diagrams likely occurred due to deviations of the actual relationship between the magnetic flux tube parameters from the simplified scaling law (\ref{Bscale}). Another factor that could contribute to splitting or broadening of the idealized degeneracy stripes were possible variations of the power-law indices $a$ and $b$ and/or the factor $Q_0$ in  heating law (\ref{HeatingRate}) within an active region. We remind here that in some of the proposed theoretical coronal heating mechanisms (see Section \ref{HeatingMechanisms}), the indices $a$ and $b$ are not fixed, but depend on some free parameters such as the spectral characteristics of the sub-photospheric driver, which, together with the driver's power, can be spatially dependent, too. Nevertheless, according to our observations and modeling, noticeable deviations from  scaling laws (\ref{HeatingRate}) and (\ref{Bscale}) in active regions occur rarely: they were detected in $\lesssim 12\%$ of cases only.

\section{Discussion}\label{Discussion}
\subsection{Scaling laws in the solar active regions}\label{Discussion_delta}
The relationships between different parameters of the coronal magnetic flux tubes in solar active regions have been analyzed before. In particular, \citet{Klimchuk1995}, using the Yohkoh/SXT soft X-ray observations, estimated the power-law index $b_0=-\alpha$ (that characterizes the effective dependence of the volumetric heating rate $Q$ on the magnetic flux tube length $L$) as $0.95\le b_0\le 3.11$ with the most probable value of $b_0=1.95$, which is consistent with the values obtained in this study (see Figure \ref{FigB0histograms}). By assuming the plasma heating mechanism based on the dissipation of stressed magnetic fields and proposed by \citet{Parker1983, Parker1988}, \citet{Klimchuk1995} estimated the corresponding power-law index $\delta$ (that characterizes the dependence of the average magnetic field strength in a magnetic flux tube $\left<B\right>$ on the length of that tube $L$) as $-1.1\le\delta\le 0.0$ with the most probable value of $\delta\simeq -0.5$, which agrees with both the most probable value and the obtained range of $\delta$ indices in this study (see Figure \ref{FigChistograms}). \citet{Klimchuk1995} also combined their results with the Skylab/S-054 soft X-ray observations reported by \citet{Golub1980} and obtained $\delta\simeq -0.7$, which agrees with the second peak in the distribution of $\delta$ in Figure \ref{FigChistograms}. \citet{Mandrini2000} and \citet{Jain2006}, using the coronal magnetic fields extrapolated from photospheric magnetograms, estimated the index $\delta$ as $\delta=-0.88\pm 0.3$ and $\delta=-1.17\pm 0.46$, respectively, which covers the region of large (by absolute value) $\delta$ indices in Figure \ref{FigChistograms}. \citet{Aschwanden1999}, using the SOHO/EIT EUV observations and SOHO/MDI magnetograms, estimated the index $\delta$ as $\delta=-1.02\pm 0.43$, which again is consistent with the large-negative tail of the distribution in Figure \ref{FigChistograms}. \citet{MacCormack2020}, using the STEREO/EUVI EUV observations and the extrapolated coronal magnetic fields, found for the \textit{quiet-Sun coronal loops} the indices $\delta$ in the range of about $-0.55\le\delta\le -0.15$, which overlaps with a significant fraction of our results.

We note that, in contrast to the above cited works, the microwave observations have allowed us not only to estimate some characteristic values of $\delta$ and $b_0$, but to reconstruct statistical distributions of those indices. Different values of $\delta$ reported by different authors, as well as broad confidence ranges, are likely caused by the fact that the mentioned index can vary significantly from one active region to another, and also with time. Additionally, the scaling-law indices inferred from the microwave, EUV, or soft X-ray observations may refer to different subsets of magnetic loops, and be subject to different weightings. Potentially, comparing the indices inferred from different observational data for the same active region can provide valuable clues about the magnetic field and plasma structures there.

\begin{figure}
\centerline{\includegraphics{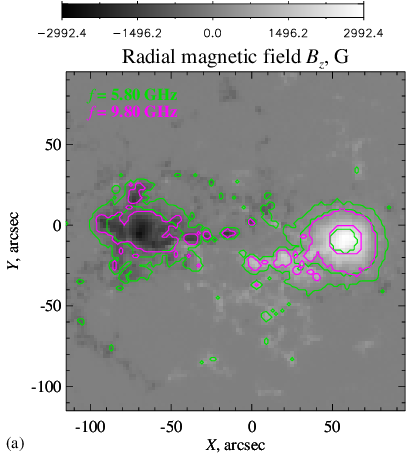}}
\centerline{\includegraphics{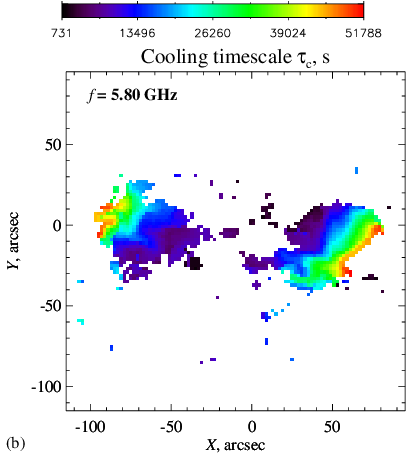}}
\caption{(a) \textit{SDO}/HMI magnetogram of AR 12936 at 2022-01-31 04:20 UT, in the top-view projection, with the reference point corresponding to the helioprojective coordinates of $(170'', 390'')$. The overplotted contours show the heights of the third-harmonic gyrolayers (for the emission frequencies of 5.80 and 9.80 GHz) above the photosphere; the contours are drawn at the levels of 0, 6000, and 12\,000 km. (b) The corresponding 2D distribution of the plasma cooling timescales $\tau_{\mathrm{c}}$ at the third-harmonic gyrolayer for the emission frequency of 5.80 GHz, for the EBTEL stochastic nanoflare heating regime with $\tau=0.2\tau_{\mathrm{c}}$ and $\alpha=-2.5$.}
\label{FigCooling}
\end{figure}

\subsection{Constraints on the nanoflare cadences}\label{NanoflareCadences}
We now estimate the values of the nanoflare cadences in physical units and compare them in different heating models. We note that the steady-state heating is actually equivalent (indistinguishable) from heating by very frequent nanoflares with the cadence much shorter than the plasma cooling time.

In the EBTEL stochastic nanoflare heating models, the median nanoflare cadences are assumed to be proportional to the local plasma cooling times $\tau_{\mathrm{c}}$ in coronal magnetic loops, i.e., they  vary within the active region. To estimate these cooling times, we selected the model volume elements located at a certain gyrolayer (we considered the third-harmonic gyrolayer, which typically makes the dominant contribution to the observed microwave emission); the shapes and heights of the gyrolayers for the emission frequencies of 5.80 and 9.80 GHz in AR 12936 are shown in Figure \ref{FigCooling}a. Then, for the magnetic field lines crossing the selected volume elements, we computed the heating rates $Q$ using the best-fit heating model parameters $(Q_0, a, b)$ inferred from the microwave observations, and estimated the average plasma densities and temperatures as the moments of the DEM distributions from the corresponding EBTEL lookup table. We note that the parameters of the magnetic flux tubes not crossing the gyrolayers for $f=6-12$\,GHz are not constrained by our observations and simulations. Knowing the magnetic loop lengths and plasma densities and temperatures, we estimate the corresponding cooling times $\tau_{\mathrm{c}}$; we used the formula (A2) by \citet{Cargill2014} with the radiative loss function in the form by \citet{Klimchuk2008}, see Appendix \ref{CoolingTimeFormulae}. An example of the obtained 2D distribution of the cooling time $\tau_{\mathrm{c}}$ over an isogauss surface within an active region is shown in Figure \ref{FigCooling}b. One can see that the cooling times are shorter near the center of the active region, and increase in the direction towards the edges of the active region, where the magnetic loops are longer. The cooling times vary in a broad range; the most typical (median) value in the case shown in Figure \ref{FigCooling}b is $\tau_{\mathrm{c}}\simeq 14\,500$ s, which for the considered there EBTEL model with $\tau=0.2\tau_{\mathrm{c}}$ corresponds to the median nanoflare cadence of $\tau\simeq 2900$ s.

\begin{table}
\caption{Typical (median) nanoflare cadences $\tau$ with the corresponding $1\sigma$ (68\%) confidence intervals, in units of $10^4$ s, derived for different active regions and EBTEL heating regimes, at the selected frequencies.}
\label{TabCadences}
\movetableright=-1.0cm
\begin{tabular}{cCCC}
\hline\hline
Model & \sim 6~\textrm{GHz} & \sim 8~\textrm{GHz} & \sim 10~\textrm{GHz}\\
\hline
 & \multicolumn{3}{c}{AR12924, 2022-01-09 04:45}\\
$\tau{=}0.2\tau_{\mathrm{c}}$, $\alpha{=}{-}2.5$ &  0.50_{ 0.25}^{ 1.86} &  0.52_{ 0.27}^{ 2.07} &  0.81_{ 0.30}^{ 2.29}\\
$\tau{=}\tau_{\mathrm{c}}$, $\alpha{=}{-}2.5$ &  2.26_{ 1.19}^{ 4.11} &  2.43_{ 1.36}^{ 9.76} &  2.87_{ 1.52}^{12.45}\\
$\tau{=}5\tau_{\mathrm{c}}$, $\alpha{=}{-}2.5$ & 11.32_{ 6.58}^{19.30} & 12.34_{ 7.28}^{20.42} & 13.32_{ 7.86}^{79.92}\\
 & \multicolumn{3}{c}{AR12936, 2022-01-31 04:20}\\
$\tau{=}0.2\tau_{\mathrm{c}}$, $\alpha{=}{-}2.5$ &  0.29_{ 0.14}^{ 0.63} &  0.31_{ 0.14}^{ 0.68} &  0.33_{ 0.16}^{ 0.72}\\
$\tau{=}\tau_{\mathrm{c}}$, $\alpha{=}{-}2.5$ &  1.45_{ 0.69}^{ 3.14} &  1.53_{ 0.72}^{ 3.41} &  1.64_{ 0.81}^{ 3.59}\\
$\tau{=}5\tau_{\mathrm{c}}$, $\alpha{=}{-}2.5$ &  7.33_{ 3.51}^{15.69} &  7.65_{ 3.64}^{17.06} &  8.20_{ 4.05}^{17.95}\\
 & \multicolumn{3}{c}{AR13007, 2022-05-14 03:25}\\
$\tau{=}0.2\tau_{\mathrm{c}}$, $\alpha{=}{-}2.5$ &  0.19_{ 0.10}^{ 0.49} &  0.20_{ 0.10}^{ 0.47} &  0.21_{ 0.10}^{ 0.46}\\
$\tau{=}\tau_{\mathrm{c}}$, $\alpha{=}{-}2.5$ &  0.98_{ 0.48}^{ 2.45} &  1.00_{ 0.49}^{ 2.35} &  1.06_{ 0.52}^{ 2.29}\\
$\tau{=}5\tau_{\mathrm{c}}$, $\alpha{=}{-}2.5$ &  5.09_{ 2.48}^{12.69} &  5.09_{ 2.48}^{12.33} &  5.57_{ 2.77}^{12.69}\\
 & \multicolumn{3}{c}{AR13234, 2023-02-26 03:40}\\
$\tau{=}0.2\tau_{\mathrm{c}}$, $\alpha{=}{-}2.5$ &  0.31_{ 0.05}^{ 0.74} &  0.33_{ 0.07}^{ 0.75} &  0.38_{ 0.14}^{ 0.77}\\
$\tau{=}\tau_{\mathrm{c}}$, $\alpha{=}{-}2.5$ &  1.54_{ 0.26}^{ 3.68} &  1.64_{ 0.36}^{ 3.75} &  1.89_{ 0.72}^{ 3.84}\\
$\tau{=}5\tau_{\mathrm{c}}$, $\alpha{=}{-}2.5$ &  7.70_{ 1.50}^{18.41} &  8.31_{ 2.25}^{19.04} &  9.47_{ 3.58}^{19.21}\\
 & \multicolumn{3}{c}{AR13245, 2023-03-09 04:00}\\
$\tau{=}0.2\tau_{\mathrm{c}}$, $\alpha{=}{-}2.5$ &  0.25_{ 0.03}^{ 0.64} &  0.33_{ 0.04}^{ 0.69} &  0.38_{ 0.10}^{ 0.82}\\
$\tau{=}\tau_{\mathrm{c}}$, $\alpha{=}{-}2.5$ &  1.27_{ 0.15}^{ 3.18} &  1.63_{ 0.21}^{ 3.44} &  1.92_{ 0.52}^{ 4.08}\\
$\tau{=}5\tau_{\mathrm{c}}$, $\alpha{=}{-}2.5$ &  6.35_{ 0.76}^{15.90} &  8.14_{ 1.18}^{17.18} &  9.88_{ 2.62}^{22.40}\\
 & \multicolumn{3}{c}{AR13315, 2023-05-28 03:55}\\
$\tau{=}0.2\tau_{\mathrm{c}}$, $\alpha{=}{-}2.5$ &  0.27_{ 0.09}^{ 0.64} &  0.29_{ 0.10}^{ 0.66} &  0.32_{ 0.14}^{ 0.70}\\
$\tau{=}\tau_{\mathrm{c}}$, $\alpha{=}{-}2.5$ &  1.35_{ 0.47}^{ 3.18} &  1.45_{ 0.52}^{ 3.28} &  1.60_{ 0.69}^{ 3.48}\\
$\tau{=}5\tau_{\mathrm{c}}$, $\alpha{=}{-}2.5$ &  6.83_{ 2.68}^{16.13} &  7.24_{ 2.62}^{16.38} &  8.02_{ 3.45}^{17.40}\\
 & \multicolumn{3}{c}{AR13372, 2023-07-17 03:45}\\
$\tau{=}0.2\tau_{\mathrm{c}}$, $\alpha{=}{-}2.5$ &  0.42_{ 0.14}^{ 0.79} &  0.49_{ 0.18}^{ 0.90} &  0.53_{ 0.27}^{ 0.91}\\
$\tau{=}\tau_{\mathrm{c}}$, $\alpha{=}{-}2.5$ &  2.11_{ 0.72}^{ 3.94} &  2.54_{ 0.89}^{ 4.53} &  2.66_{ 1.34}^{ 4.51}\\
$\tau{=}5\tau_{\mathrm{c}}$, $\alpha{=}{-}2.5$ & 10.77_{ 3.65}^{20.07} & 12.29_{ 4.39}^{22.18} & 13.26_{ 6.68}^{22.10}\\
 & \multicolumn{3}{c}{AR13559, 2024-01-24 04:15}\\
$\tau{=}0.2\tau_{\mathrm{c}}$, $\alpha{=}{-}2.5$ &  0.33_{ 0.08}^{ 0.58} &  0.36_{ 0.09}^{ 0.59} &  0.41_{ 0.13}^{ 0.62}\\
$\tau{=}\tau_{\mathrm{c}}$, $\alpha{=}{-}2.5$ &  1.65_{ 0.41}^{ 2.90} &  1.78_{ 0.43}^{ 2.96} &  2.04_{ 0.64}^{ 3.11}\\
$\tau{=}5\tau_{\mathrm{c}}$, $\alpha{=}{-}2.5$ &  8.28_{ 2.28}^{14.52} &  8.91_{ 2.22}^{14.80} & 10.20_{ 3.20}^{15.55}\\
\hline
 & \multicolumn{3}{c}{Combined}\\
$\tau{=}0.2\tau_{\mathrm{c}}$, $\alpha{=}{-}2.5$ &  0.32_{ 0.10}^{ 0.68} &  0.34_{ 0.11}^{ 0.74} &  0.39_{ 0.15}^{ 0.80}\\
$\tau{=}\tau_{\mathrm{c}}$, $\alpha{=}{-}2.5$ &  1.58_{ 0.51}^{ 3.33} &  1.71_{ 0.56}^{ 3.61} &  1.93_{ 0.75}^{ 3.95}\\
$\tau{=}5\tau_{\mathrm{c}}$, $\alpha{=}{-}2.5$ &  7.95_{ 2.67}^{16.80} &  8.61_{ 2.96}^{17.87} &  9.76_{ 3.90}^{18.69}\\
\hline
\end{tabular}
\end{table}

In Table \ref{TabCadences}, we present the estimated median nanoflare cadences for all eight considered active regions, at the same times and frequencies as in Table \ref{TabMultiEBTEL}. The stochastic nanoflare heating models with $\alpha=-2.5$ and different $\tau/\tau_{\mathrm{c}}$ scaling factors are compared; for the models with $\alpha=-1$, the results have been found to be almost the same. One can see that the median nanoflare cadences tend to increase slightly with the emission frequency, which occurs because at higher frequencies, the gyrolayers are located predominantly in the regions with longer magnetic loops and hence longer cooling times (see Figure \ref{FigCooling}). 

For the high-frequency stochastic heating model (with $\tau=0.2\tau_{\mathrm{c}}$), the typical nanoflare cadences are of about $1-2$ hours. As demonstrated above, the high-frequency and steady-state heating regimes provided very similar results, although the steady-state heating was typically marginally better in reproducing the observations. Thus we conclude that in the cases when the steady-state EBTEL heating regime has been found to be the most favourable one (i.e., in the majority of the considered events), the plasma heating was either truly continuous (e.g., wave heating) or caused by nanoflares with a cadence shorter than $\sim 1$ hour.

For the intermediate-frequency stochastic heating model (with $\tau=\tau_{\mathrm{c}}$), the typical nanoflare cadences are of about $(1-2)\times 10^4$ s or $3-5$ hours, which is close to the fixed nanoflare cadence of $10^4$ s in the impulsive heating model. We remind that the intermediate-frequency stochastic heating models and the impulsive heating model have been found to provide comparable model-to-observations fitting quality, which can be attributed to comparable nanoflare cadences, although the particular best-fit $a$ and $b$ values were somewhat different. In four cases, these intermediate-frequency heating regimes have been found to be the most favourable ones.

For the low-frequency stochastic heating model (with $\tau=5\tau_{\mathrm{c}}$), the typical nanoflare cadences are of about one day, which seems to be too long: as demonstrated in Section \ref{HeatingRegime}, the models with $\tau=5\tau_{\mathrm{c}}$ were always less successful in reproducing the observations than the models implying a higher-frequency heating. It is interesting to note that the impulsive heating model with a fixed cadence of $\tau=10^4$ s \citep[that was used in the earlier work by][]{Fleishman2021a} can correspond to either high-frequency ($\tau\ll\tau_{\mathrm{c}}$), intermediate-frequency ($\tau\sim\tau_{\mathrm{c}}$), or low-frequency ($\tau\gg\tau_{\mathrm{c}}$) heating regime in different parts of an active region, although the intermediate-frequency heating seems to dominate.

\begin{figure}
\centerline{\includegraphics{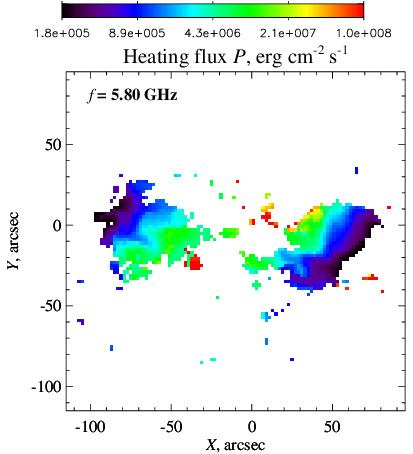}}
\caption{2D distribution of the heating flux $P$ at the third-harmonic gyrolayer for the emission frequency of 5.80 GHz, in AR 12936 at 2022-01-31 04:20 UT (the gyrolayer structure is shown in Figure \protect\ref{FigCooling}a), for the steady-state EBTEL heating regime.}
\label{FigHeatingFlux}
\end{figure}

\subsection{Heating fluxes}\label{HeatingFluxes}
For a magnetic flux tube with the length $L$ and volumetric heating rate $Q$, the corresponding heating flux per unit area $P$ (measured in erg $\textrm{cm}^{-2}$ $\textrm{s}^{-1}$) can be estimated as $P=QL$. Figure \ref{FigHeatingFlux} demonstrates an example of the model-derived 2D distribution of the heating flux $P$ over an isogauss surface within an active region; the heating flux values for the magnetic flux tubes crossing a selected gyrolayer were computed following the approach described in previous Section. The heating fluxes are the highest near the center of the active region, and tend to decrease towards the edges of the active region. This occurs because the heating rate is scaled with the magnetic flux tube length as $Q\propto L^{-b}$, so that for the heating flux we have $P\propto L^{1-b}$, and for $b\simeq 2.7>1$ (as in the case shown in Figure \ref{FigHeatingFlux}) the heating flux is higher for shorter loops, i.e., closer to the center of the active region. The heating fluxes vary in a broad range (by almost three orders of magnitude), while the most typical (median) value is $P\simeq 1.72\times 10^6$ erg $\textrm{cm}^{-2}$ $\textrm{s}^{-1}$.

\begin{table}
\caption{Typical (median) heating fluxes $P$ with the corresponding $1\sigma$ (68\%) confidence intervals, in units of $10^6$ erg $\textrm{cm}^{-2}$ $\textrm{s}^{-1}$, derived for different active regions at different times, for the steady-state EBTEL heating regime, at the selected frequencies.}
\label{TabHeatingFluxes}
\movetableright=-1.0cm
\begin{tabular}{cCCC}
\hline\hline
Time, UT & \sim 6~\textrm{GHz} & \sim 8~\textrm{GHz} & \sim 10~\textrm{GHz}\\
\hline
2022-01-09 04:45 &  0.60_{ 0.11}^{ 2.88} &  0.43_{ 0.08}^{ 1.94} &  0.27_{ 0.07}^{ 1.18}\\
2022-01-09 05:35 &  0.53_{ 0.09}^{ 2.93} &  0.53_{ 0.13}^{ 2.26} &  0.36_{ 0.12}^{ 1.25}\\
2022-01-10 05:25 &  0.85_{ 0.24}^{ 2.09} &  0.54_{ 0.31}^{ 0.96} &  0.38_{ 0.29}^{ 0.43}\\
2022-01-10 06:10 &  0.72_{ 0.22}^{ 2.12} &  0.49_{ 0.26}^{ 0.97} &  0.43_{ 0.25}^{ 0.76}\\
2022-01-30 04:15 &  5.68_{ 3.78}^{ 8.10} &  4.83_{ 3.21}^{ 6.82} &  5.53_{ 3.42}^{ 7.71}\\
2022-01-30 06:25 &  3.59_{ 3.07}^{ 5.09} &  2.20_{ 1.74}^{ 4.43} &  2.67_{ 2.11}^{ 4.63}\\
2022-01-31 04:20 &  1.72_{ 0.37}^{ 7.31} &  1.62_{ 0.53}^{ 4.76} &  1.46_{ 0.52}^{ 3.83}\\
2022-01-31 06:15 &  2.25_{ 0.49}^{ 8.99} &  1.63_{ 0.57}^{ 4.73} &  1.35_{ 0.46}^{ 4.45}\\
2022-05-14 03:25 &  4.64_{ 1.26}^{12.54} &  4.47_{ 2.67}^{ 7.40} &  3.94_{ 1.79}^{ 8.15}\\
2022-05-14 06:35 &  4.63_{ 1.16}^{11.08} & \nodata & \nodata\\
2022-05-15 03:45 &  4.51_{ 1.41}^{12.60} &  3.53_{ 1.39}^{ 8.97} &  3.03_{ 1.13}^{ 6.77}\\
2022-05-15 06:20 &  4.80_{ 1.14}^{21.66} & \nodata &  3.11_{ 0.18}^{14.60}\\
2023-02-26 03:40 &  1.65_{ 0.45}^{24.33} &  1.30_{ 0.39}^{13.63} &  0.96_{ 0.30}^{ 4.64}\\
2023-02-26 06:40 &  3.18_{ 1.02}^{21.12} &  2.01_{ 0.69}^{17.03} &  1.32_{ 0.47}^{ 5.71}\\
2023-02-27 03:45 &  1.51_{ 0.50}^{16.49} &  1.19_{ 0.37}^{14.15} &  0.88_{ 0.27}^{ 9.75}\\
2023-02-27 06:45 &  1.98_{ 0.64}^{25.69} &  1.19_{ 0.46}^{12.93} &  0.83_{ 0.28}^{11.91}\\
2023-03-09 04:00 &  2.54_{ 0.87}^{31.41} &  2.11_{ 0.87}^{18.44} &  1.26_{ 0.48}^{ 6.41}\\
2023-03-09 06:25 &  4.74_{ 1.75}^{39.92} &  2.21_{ 0.92}^{20.64} &  1.11_{ 0.16}^{11.51}\\
2023-03-10 03:45 &  5.47_{ 2.17}^{31.26} &  3.87_{ 2.05}^{17.95} &  2.59_{ 1.41}^{11.93}\\
2023-03-10 06:20 &  5.54_{ 1.80}^{61.56} &  3.99_{ 2.41}^{15.96} &  2.86_{ 1.49}^{ 7.65}\\
2023-05-27 03:50 &  2.18_{ 0.25}^{20.30} & \nodata &  2.55_{ 0.48}^{15.03}\\
2023-05-27 06:30 &  2.60_{ 0.21}^{22.31} &  1.85_{ 0.20}^{17.74} &  2.35_{ 0.25}^{17.47}\\
2023-05-28 03:55 &  1.65_{ 0.46}^{10.82} &  1.73_{ 0.65}^{ 7.45} &  1.71_{ 0.67}^{ 5.34}\\
2023-05-28 06:50 &  2.41_{ 0.62}^{12.07} &  2.39_{ 0.81}^{ 7.72} &  2.33_{ 0.83}^{ 6.54}\\
2023-07-17 03:45 &  1.04_{ 0.47}^{ 6.09} &  0.82_{ 0.21}^{ 2.73} &  0.52_{ 0.13}^{ 1.20}\\
2023-07-17 06:15 &  1.18_{ 0.62}^{ 7.24} &  0.73_{ 0.39}^{ 3.92} &  0.57_{ 0.21}^{ 1.66}\\
2023-07-18 04:00 &  1.73_{ 0.59}^{ 8.20} &  1.36_{ 0.79}^{ 4.96} &  0.91_{ 0.69}^{ 2.56}\\
2023-07-18 06:40 &  1.77_{ 0.58}^{ 9.73} &  1.41_{ 0.56}^{ 7.25} &  0.78_{ 0.47}^{ 3.09}\\
2024-01-23 04:20 &  2.32_{ 0.98}^{27.31} &  1.84_{ 0.82}^{21.28} &  2.14_{ 1.07}^{15.41}\\
2024-01-23 05:50 &  2.69_{ 1.21}^{23.14} &  2.36_{ 1.12}^{25.87} &  2.75_{ 0.37}^{17.21}\\
2024-01-24 04:15 &  1.79_{ 0.77}^{14.30} &  1.21_{ 0.53}^{11.99} &  0.95_{ 0.53}^{ 5.29}\\
2024-01-24 06:15 &  1.31_{ 0.54}^{11.19} &  1.06_{ 0.46}^{ 8.68} &  0.79_{ 0.35}^{ 3.89}\\
\hline
Combined &  2.41_{ 0.61}^{12.91} &  1.75_{ 0.53}^{ 7.82} &  1.37_{ 0.39}^{ 6.53}\\
\hline
\end{tabular}
\end{table}

In Table \ref{TabHeatingFluxes}, we present the estimated heating fluxes for all considered active regions and times, at three selected frequencies, for the steady-state EBTEL heating regime. The median heating fluxes tend to decrease with an increase of the emission frequency, which is caused by the above mentioned changes in the geometry of the gyrolayers that tend to encompass predominantly longer magnetic loops (with lower heating fluxes) for higher frequencies. The median heating fluxes for all considered active regions combined were of about $(1-2)\times 10^6$ erg $\textrm{cm}^{-2}$ $\textrm{s}^{-1}$, although in individual active regions they varied from $\sim 3\times 10^5$ to $\sim 6\times 10^6$ erg $\textrm{cm}^{-2}$ $\textrm{s}^{-1}$. These values agree well with the previously estimated coronal heating fluxes in active regions \citep[$\sim 10^6-10^7$ erg $\textrm{cm}^{-2}$ $\textrm{s}^{-1}$, according to][]{Withbroe1977, Aschwanden2001}. At the same time, our simulations predict that in some (short) magnetic loops the heating fluxes can be much higher---up to $\sim 10^8$ erg $\textrm{cm}^{-2}$ $\textrm{s}^{-1}$, but these overheated magnetic loops, due to their small sizes, are unlikely to make a significant contribution to the microwave emission.

\begin{figure}
\centerline{\includegraphics{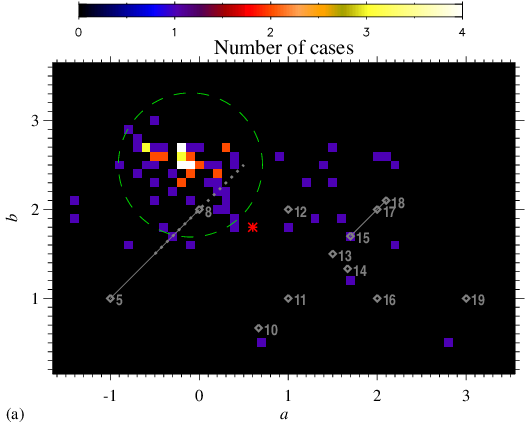}}
\centerline{\includegraphics{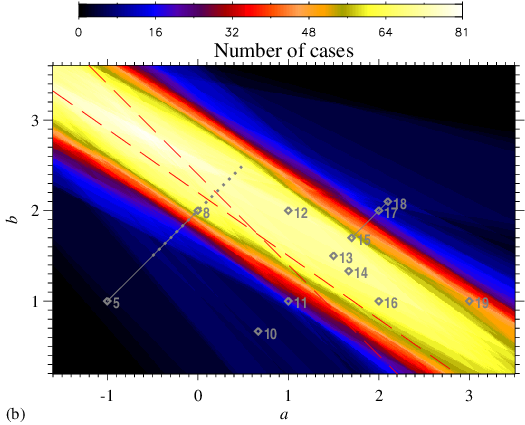}}
\caption{(a) 2D histogram of the best-fit parameters $(a, b)$ of the coronal heating model, with the bin size of 0.1 in both directions; the green dashed circle bounds an area where 68\% of all best-fit $(a, b)$ combinations are located. (b) ``Combined'' $a{-}b$ diagram demonstrating how many times a given $(a, b)$ combination fitted the SRH observations, assuming the degeneracy stripes of infinite length and with the width of $w=1$. All results correspond to the steady-state EBTEL heating regime. Additionally, the red asterisk in panel (a) and dashed red lines in panel (b) show respectively the best-fit $(a, b)$ combination and the axes of the degeneracy stripes for AR 11520 at the frequencies of 5.7 and 17 GHz \protect\citep{Fleishman2025}. The gray diamonds and numbers correspond to the theoretical coronal heating mechanisms presented in Table \protect\ref{TabHeatingModels}. For the models depending on a free parameter, the characteristic points are connected together by solid gray lines. The dotted gray lines correspond to the model of resonance cavities (\#5, \#8 in Table \protect\ref{TabHeatingModels}) with the index $m$ in the range of $-1.5<m<-0.5$.}
\label{FigARsummary}
\end{figure}

\subsection{Heating mechanism}\label{HeatingMechanisms}
One of the main aims of this study is to identify the actual characteristics of the coronal heating mechanism, such as the indices $a$ and $b$ in the parametric heating law (\ref{HeatingRate}). Figure \ref{FigARsummary}a is a graphical representation of the data from Table \ref{TabMultiARs}, i.e., it demonstrates a 2D distribution of the best-fit parameters $a$ and $b$ for all considered active regions at all considered times and frequencies, for the steady-state EBTEL heating regime; the regions with very large $a$ or very small $b$ are truncated for clarity. As has been said above, the most typical (median) values of $a$ and $b$ were $a\simeq -0.1$ and $b\simeq 2.5$; about 68\% ($1\sigma$) of all best-fit $(a, b)$ combinations were located within a radius of $r=0.81$ from that point. At the same time, we remind that the coronal heating model is considerably degenerate with respect to the parameters $a$ and $b$: the best-fit $a$ and $b$ values are poorly determined and can sometimes jump back and forth along the degeneracy stripe; this conclusion is confirmed by a significant scatter of dots in Figure \ref{FigARsummary}a. Statistical analysis of multiple active regions may potentially help to relieve the mentioned degeneracy, since the most likely $(a, b)$ pair is expected to be located at the intersection of all degeneracy stripes and thus correspond to the ``true'' unique combination of $a$ and $b$. However, the results presented in Figure \ref{FigARsummary}a still do not allow us to firmly constrain the heating model because the uncertainty is large. Indeed: a) the degeneracy stripes are more-or-less parallel to each other (their slopes $\delta$ are rather similar), b) the indices $a$ and $b$ could actually not be the same in different active regions, and c) the sample of active regions is limited.

\begin{table*}
\caption{Characteristics of the theoretical coronal heating mechanisms: the power-law indices in the parametric heating law $a$ and $b$, brief descriptions, and sample references. The parameter $N_{\mathrm{cases}}$ represents the number of cases (i.e., different active regions at different times) where a given theoretical model agreed with the SRH observations, at the selected frequencies and at all those frequencies in total. The $N_{\mathrm{cases}}$ numbers exceeding 75\% of the numbers of all considered cases are underlined.}
\label{TabHeatingModels}
\movetableright=-2.4cm
\renewcommand{\arraystretch}{0.975}
\begin{tabular}{cCCccccccc}
\hline\hline
 & & & & & & \multicolumn{4}{C}{N_{\mathrm{cases}}}\\
\cline{7-10}
\# & a & b & Model description & References & & ${\sim}6$ GHz & ${\sim}8$ GHz & ${\sim}10$ GHz & Total\\
\hline
       1 & -2 & 0 & Switch-on MHD shock train & 1 & &       0 &        0 &        0 &        0\\
       2 & -1 & -2 & Resonant absorption A ($m=-2$) & 2 & &       0 &        0 &        0 &        0\\
       3 & -1 & -1 & Resonant absorption B ($m=-2$) & 3, 4 & &       0 &        0 &        0 &        0\\
 & & & Hybrid triple-correlation cascade & 5 & & & & & \\			
       4 & -1 & 0 & Fast-mode shock train & 1 & &       0 &        0 &        0 &        0\\
 & & & Iroshnikov-Kraichnan cascade & 6 & & & & & \\
       5 & -1 & 1 & Resonance cavities ($m=-2$) & 1 & &       0 &        1 &        1 &        2\\
       6 & 0 & -1 & Resonant absorption A ($m=-1$) & 2 & &       0 &        0 &        0 &        0\\
       7 & 0 & 0 & Resonant absorption B ($m=-1$) & 3, 4 & &       0 &        1 &        1 &        2\\
 & & & Alfv\'en-wave collisional damping & 7 & & & & & \\	
 & & & Kolmogorov-Obukhov cascade & 8 & & & & & \\	
 & & & Reflection-driven cascade & 9 & & & & & \\				
       8 & 0 & 2 & Resonance cavities ($m=-1$) & 1 & &      19 &       \underline{23} &       \underline{20} &       62\\
       9 & 1/2 & 0 & Surface-wave damping & 1 & &       1 &        1 &        1 &        3\\
      10 & 2/3 & 2/3 & Phase mixing & 10 & &       2 &        2 &        2 &        6\\
      11 & 1 & 1 & Current layers & 11 & &       5 &        6 &        7 &       18\\
 & & & Flux cancellation & 12 & & & & & \\		
 & & & Resonant absorption B ($m=0$) & 4 & & & & & \\		
 & & & Tearing-mode reconnection & 13 & & & & & \\					
      12 & 1 & 2 & Reconnection $\propto v_{\mathrm{A}}$ & 14 &  &     \underline{26} &       \underline{25} &       \underline{22} &       \underline{73}\\
      13 & 3/2 & 3/2 & Reconnection $\propto v_{\mathrm{A}\bot}$ & 15 & &      \underline{25} &       \underline{25} &       \underline{21} &       \underline{71}\\
 & & & Turbulence with constant dissipation coefficients & 16, 17 & & & & & \\		
 & & & Line-tied reduced MHD cascade & 18 & & & & & \\					
      14 & 5/3 & 4/3 & Turbulence with closure & 19, 20, 21 & &      \underline{25} &       \underline{25} &       \underline{21} &       \underline{71}\\
 & & & Turbulence (high-frequency) & 22 & & & & & \\					
      15 & 1.7 & 1.7 & Turbulence with closure + spectrum ($m=-1$) & 23 & &      \underline{24} &       \underline{24} &       19 &       \underline{67}\\
      16 & 2 & 1 & Critical angle & 24, 25 & &      \underline{23} &       \underline{24} &       \underline{20} &       \underline{67}\\
 & & & Current sheets & 26 & & & & & \\				
      17 & 2 & 2 & Stochastic buildup & 27, 28 & &       4 &        3 &        4 &       11\\
 & & & Critical twist & 29 & & & & & \\		
 & & & Current layers & 11, 30, 31 & & & & & \\		
 & & & Taylor relaxation & 13, 32, 33 & & & & & \\					
      18 & 2.1 & 2.1 & Turbulence with closure + spectrum ($m=-2.5$) & 23 & &       4 &        2 &        3 &        9\\
      19 & 3 & 1 & Non-ideal/slipping reconnection & 34 & &      20 &       17 &       13 &       50\\
 & & & Hyperdiffusive reconnection & 35 & & & & & \\
\hline
\end{tabular}
\tablerefs{(1) \citet{Hollweg1985}; (2) \citet{Halberstadt1995}; (3) \citet{Ofman1995}; (4) \citet{Ruderman1997}; (5) \citet{Zhou1990}; (6) \citet{Chae2002}; (7) \citet{Osterbrock1961}; (8) \citet{Hollweg1986}; (9) \citet{Hossain1995}; (10) \citet{Roberts2000}; (11) \citet{Galsgaard1996}; (12) \citet{Priest2018}; (13) \citet{Browning1986}; (14) \citet{Parker1983}; (15) \citet{Parker1983}, modified; (16) \citet{Einaudi1996}; (17) \citet{Dmitruk1997}; (18) \citet{Dmitruk1999}; (19) \citet{Heyvaerts1992}; (20) \citet{Inverarity1995}; (21) \citet{Inverarity1995a}; (22) \citet{Inverarity1995b}; (23) \citet{Milano1997}; (24) \citet{Parker1988}; (25) \citet{Berger1993}; (26) \citet{Aly1997}; (27) \citet{Sturrock1981}; (28) \citet{Berger1991}; (29) \citet{Galsgaard1997}; (30) \citet{vanBallegooijen1986}; (31) \citet{Hendrix1996}; (32) \citet{Heyvaerts1984}; (33) \citet{Vekstein1993}; (34) \citet{Yang2018}; (35) \citet{vanBallegooijen2008}.}
\end{table*}

Instead, we now check which coronal heating mechanisms agree (or do not agree) with the observations, using the characteristics of the degeneracy stripes $\delta$ and $b_0$ that seem to be determined more reliably than the best-fit $a$ and $b$ indices. In Table \ref{TabHeatingModels}, we compiled a list of known theoretical coronal heating mechanisms, following the papers of \citet{Mandrini2000} and \citet{Cranmer2019}; the mechanisms are sorted by the corresponding $a$ and $b$ indices. We note that while \citet{Mandrini2000} provided the scaling laws in the form of $Q\propto\left<B\right>^aL^{-b}$, \citet{Cranmer2019} characterized the heating models by the efficiency factor $\mathcal{E}\propto\Lambda^{\alpha}\Theta^{\beta}$, where $\Lambda$ and $\Theta$ are the normalized horizontal correlation length and frequency of the sub-photospheric driver; different power-law indices are related to each other as $a=2-\beta$ and $b=\alpha-\beta+1$. Some of the mechanisms involve a continuous free parameter---the power-law spectral index $m$ of the external (sub-photospheric) driver of the magnetic field disturbances or MHD oscillations; in such cases, we considered several representative values of $m$ and the corresponding values of $a$ and $b$. The $(a, b)$ combinations for different theoretical coronal heating mechanisms are also shown in Figure \ref{FigARsummary} (some points fell beyond the range of the figure). The requirement is that the theoretically predicted $(a, b)$ combination should fall within the degeneration stripe(s).

For each of the considered active regions, at each of the considered times and at each of three selected frequencies (see Table \ref{TabMultiARs}), we identified the theoretical coronal heating mechanisms that were consistent with the observations, i.e., where the corresponding pairs of $a$ and $b$ indices were located within the degeneracy stripes in the $a{-}b$ diagrams. Consequently, we computed how many times each heating mechanism agreed with the observations. We used the ``idealized'' degeneracy stripes with infinite lengths and fixed widths of $w=1$ (that was a typical value in our observations and analysis), and with the slopes $\delta$ and intercepts $b_0$ as presented in Table \ref{TabMultiARs}. We considered only the cases when the degeneracy stripes were well-defined and narrow (namely, with the widths of $w<3$ in Table \ref{TabMultiARs}), which provided 30 cases at the frequency of $\sim 6$ GHz, 29 cases at $\sim 8$ GHz, and 26 cases at $\sim 10$ GHz, i.e., 85 cases in total. The obtained results are presented in Table \ref{TabHeatingModels}. In addition, the numbers of matches with observations for selected $(a, b)$ combinations (for all considered frequencies) are printed in the diagram in Figure \ref{FigARsummary}b.

One can see from Table \ref{TabHeatingModels} and Figure \ref{FigARsummary}b that some of the theoretical coronal heating mechanisms (e.g., \#1--4 and \#6) predict the $a$ and $b$ values that did not provide an acceptable agreement with the observations, and therefore these mechanisms can be effectively ruled out. On the other hand, among the most likely ones are the mechanisms (or groups of mechanisms) \#12--16, with $(a, b)$ in the range from $(1, 2)$ to $(2, 1)$; these mechanisms agreed with the observations in $75-85\%$ of cases.

Surprisingly, the somewhat exotic coronal heating mechanism that involves selective transmission or reflection of MHD waves in the coronal loops acting as resonance cavities (\#5, \#8), as proposed by \citet{Hollweg1985}, proved to be the best one. With $a=m+1$ and $b=m+3$, for a reasonable range of $-1.5<m<-0.5$ (with $m\simeq -0.7$ being the most likely value), this mechanism was consistent (in terms of the pairs of $a$ and $b$ being located within the degeneracy stripes) with all analyzed observations. In addition, the resonance cavities mechanism provided a reasonably good (not perfect, but the best among all considered mechanisms) agreement with the particular best-fit $a$ and $b$ values, as can be seen in Figure \ref{FigARsummary}a. This mechanism is also naturally consistent with the steady-state heating regime that has been found to be the most likely one (see Section \ref{HeatingRegime}). Although the resonance cavities model by \citet{Hollweg1985} is too general and lacks many important details, that model may reflect adequately certain processes in the solar corona. Other theoretical coronal heating mechanisms involving the free parameter $m$ (based on resonant absorption or turbulence with closure + spectrum, \#2--3, \#6--7, \#11, \#15, \#18), although consistent with the observations in some cases, were unable to provide an agreement with all analyzed observations for reasonable values of $m$.

For comparison, we also showed in Figure \ref{FigARsummary} the best-fit coronal heating model parameters that have been obtained by \citet{Fleishman2025} for AR 11520 on 2012-07-12, using the observations of the Siberian Solar Radio Telescope and Nobeyama Radioheliograph at the frequencies of 5.7 and 17 GHz, respectively. In that case, the observations were best reproduced by the high-frequency heating model with $\tau=0.2\tau_{\mathrm{c}}$ and $\alpha=-2.5$, the best-fit heating law indices were $a\simeq 0.6$ and $b\simeq 1.8$ at both frequencies; the degeneracy stripes at the two frequencies were slightly different but intersected approximately at the mentioned best-fit $(a, b)$ solution. The results of \citet{Fleishman2025} are well in line with the results obtained in this study, although a broader frequency range allowed for a better refinement of the heating model parameters in AR 11520.

\section{Conclusion}
We analyzed the microwave observations of eight solar active regions obtained with the Siberian Radioheliograph in years $2022-2024$ in the frequency range of $\sim 6-12$ GHz. We performed forward-fitting modeling of the gyroresonance microwave emission from the considered active regions, using the GX Simulator tool and the dedicated Coronal Heating Modeling Pipeline code. The aim was to determine the most likely characteristics of the  heating mechanism of the diffuse component of the coronal plasma, including the power-law indices $a$ and $b$ that determine the dependences of the local volumetric heating rate $Q$ respectively on the average magnetic field strength in a magnetic flux tube $\left<B\right>$ and on the length of that tube $L$, in the form of $Q\propto\left<B\right>^aL^{-b}$. We have found that:

-- Determining the exact parameters of the heating mechanism is difficult, because the average magnetic field strengths and lengths of the magnetic field tubes in solar active regions are mutually dependent, which makes the coronal heating model degenerate with respect to $a$ and $b$. Instead, the solutions of the forward-fitting problem form narrow stripes (``degeneracy stripes'') in the $(a, b)$ space. The slopes and locations of those stripes reflect the relationships between different physical quantities in the active regions ($\left<B\right>$ vs. $L$ and $Q$ vs. $L$, respectively).

-- In the considered range of the emission frequencies ($\sim 6-12$ GHz), the inferred parameters of the coronal heating model demonstrated no significant variations with the frequency. This result indicates that the basic scaling laws describing the relationships between different physical quantities were mostly the same within each of the considered active regions, and did not vary significantly with height.

-- For the relationship between the average magnetic field strength in a magnetic tube and the length of that tube, in the form of $\left<B\right>\propto L^{\delta}$, the estimated indices $\delta$ were in the range from $-1.2$ to $-0.2$, with the most probable value of $\delta\simeq -0.55$. For the effective dependence of the volumetric heating rate on the length of a magnetic flux tube, in the form of $Q\propto L^{-b_0}$, the estimated indices $b_0$ were in the range from 0.3 to 3.2, with the most probable value of $b_0\simeq 2.5$. These indices varied significantly from one active region to another; in a chosen active region, they did not demonstrate significant variations on the timescales of a few hours, but could change noticeably within one day.

-- In most cases, the steady-state (continuous) heating regime provided the best agreement of the models with the observations, which favours either the wave heating or the high-frequency heating by nanoflares with cadences much shorter ($\lesssim 1$ hour) than the plasma cooling times in the coronal magnetic loops. This conclusion supports the recent findings of \citet{Fleishman2025}, but contradicts those of, e.g., \citet{Viall2017, Mondal2024}. 

-- In a few active regions at certain frequencies, intermediate-frequency heating by nanoflares with cadences comparable with the plasma cooling times of $\sim 10^4$ s (but not much longer) appeared to be a preferable heating regime. Such cases might be related to selective heating of individual hot magnetic loops; the selective heating will be analyzed in detail in a separate study.

-- The most likely values of the indices $a$ and $b$ in the parametric heating law were $a\simeq -0.1$ and $b\simeq 2.5$. Thus in the corresponding (i.e., consistent with the majority of the observations) coronal heating models the volumetric heating rates were nearly independent on the average magnetic field strengths in magnetic loops, and only the dependence on the loop lengths with the power-law index of $b\simeq b_0\simeq 2.5$ remained. At the same time, the particular best-fit values of the indices $a$ and $b$ demonstrated a significant scatter from one considered event to another, varying, at least, in the ranges of $-1.4\le a\le 8.9$ and $-5.6\le b\le 3.0$.

-- The inferred coronal heating fluxes varied in a broad range within the considered active regions, with the most typical values of about $(1-2)\times 10^6$ erg $\textrm{cm}^{-2}$ $\textrm{s}^{-1}$. The inferred total heating rates were of the order of $\sim 10^{25}-10^{26}$ erg $\textrm{s}^{-1}$ per active region and demonstrated a positive correlation with the measured microwave fluxes.

-- Among the known theoretical coronal heating mechanisms, the model of wave heating in resonance cavities by \citet{Hollweg1985} has been found to provide the best agreement with the observations. Other models consistent with the majority of the observations (albeit with larger deviations from the best-fit $a$ and $b$ indices) were those with $(a, b)$ in the range from $(1, 2)$ to $(2, 1)$, i.e., models \#12-16 in Table \ref{TabHeatingModels}, proposed at different times by \citet{Parker1983, Parker1988, Einaudi1996, Dmitruk1997, Dmitruk1999, Heyvaerts1992, Inverarity1995, Inverarity1995a, Inverarity1995b, Milano1997, Berger1993, Aly1997}.

The magneto-thermal models of active regions used in this study are oversimplified, as they account only for the average (diffuse) component of the coronal plasma but do not account for isolated selectively heated magnetic loops (which look like bright loops in EUV images) and open magnetic loops (where the plasma heating mechanism can be different); \citep[see discussion in][]{Fleishman2025}. Also, the used EBTEL tables were computed for uniform cylindrical flux tubes, thus ignoring the loop expansion in the corona. Another source of uncertainty is the possible deviations of the actual magnetic field structure from the used nonlinear force-free extrapolation. Nevertheless, the performed simulations provided a very good agreement with the observed microwave images---both visual and quantified by the residual. As demonstrated by \citet{Fleishman2025}, the parameters of the coronal heating model inferred from the microwave observations allow one to reproduce successfully the diffuse (i.e., excluding the selectively heated bright loops) coronal EUV emission of active regions. Thus we believe that the obtained constraints on the coronal heating model are reliable and can be used as a solid basis for further development of the theoretical models of coronal heating.

\begin{acknowledgments}
This work was supported by the Ministry of Science and Higher Education of the Russian Federation (AAK and SAA) as well as supported in part by NSF grants RISE-2324724 and AST-2206424 and NASA grant 80NSSC23K0090 to New Jersey Institute of Technology (GDF and GMN). We thank the team of the Siberian Radioheliograph (unique research facility identifier \url{https://ckp-rf.ru/catalog/usu/4138190/}) for maintaining the instrument.
\end{acknowledgments}

\bibliographystyle{aasjournalv7}
\bibliography{CoronalHeating}

\begin{thebibliography}{}
\expandafter\ifx\csname natexlab\endcsname\relax\def\natexlab#1{#1}\fi
\providecommand{\url}[1]{\href{#1}{#1}}
\providecommand{\dodoi}[1]{doi:~\href{http://doi.org/#1}{\nolinkurl{#1}}}
\providecommand{\doeprint}[1]{\href{http://ascl.net/#1}{\nolinkurl{http://ascl.net/#1}}}
\providecommand{\doarXiv}[1]{\href{https://arxiv.org/abs/#1}{\nolinkurl{https://arxiv.org/abs/#1}}}

\bibitem[{C.~E. {Alissandrakis} \& D.~E. {Gary}(2021){Alissandrakis} \&
  {Gary}}]{Alissandrakis2021}
{Alissandrakis}, C.~E., \& {Gary}, D.~E. 2021, \bibinfo{title}{{Radio
  Measurements of the Magnetic field in the Solar Chromosphere and the
  Corona},} Frontiers in Astronomy and Space Sciences, 7, 77,
  \dodoi{10.3389/fspas.2020.591075}

\bibitem[{A. {Altyntsev} {et~al.}(2020){Altyntsev}, {Lesovoi}, {Globa},
  {Gubin}, {Kochanov}, {Grechnev}, {Ivanov}, {Kobets}, {Meshalkina}, {Muratov},
  {Prosovetsky}, {Myshyakov}, {Uralov}, \& {Fedotova}}]{Altyntsev2020}
{Altyntsev}, A., {Lesovoi}, S., {Globa}, M., {et~al.} 2020,
  \bibinfo{title}{{Multiwave Siberian Radioheliograph},} Solar-Terrestrial
  Phys., 6, 30, \dodoi{10.12737/stp-6220200310.12737/szf-62202003}

\bibitem[{J.~J. {Aly} \& T. {Amari}(1997){Aly} \& {Amari}}]{Aly1997}
{Aly}, J.~J., \& {Amari}, T. 1997, \bibinfo{title}{{Current sheets in
  two-dimensional potential magnetic fields. III. Formation in complex topology
  configurations and application to coronal heating.},} \aap, 319, 699

\bibitem[{I. {Arregui} \& T. {Van Doorsselaere}(2024){Arregui} \& {Van
  Doorsselaere}}]{Arregui2024}
{Arregui}, I., \& {Van Doorsselaere}, T. 2024, in Magnetohydrodynamic Processes
  in Solar Plasmas, ed. A.~K. {Srivastava}, M.~{Goossens}, \& I.~{Arregui}
  (Elsevier), 415--450, \dodoi{10.1016/B978-0-32-395664-2.00015-3}

\bibitem[{M.~J. {Aschwanden}(2001){Aschwanden}}]{Aschwanden2001}
{Aschwanden}, M.~J. 2001, \bibinfo{title}{{An Evaluation of Coronal Heating
  Models for Active Regions Based on Yohkoh, SOHO, and TRACE Observations},}
  \apj, 560, 1035, \dodoi{10.1086/323064}

\bibitem[{M.~J. {Aschwanden} {et~al.}(1999){Aschwanden}, {Newmark},
  {Delaboudini{\`e}re}, {Neupert}, {Klimchuk}, {Gary}, {Portier-Fozzani}, \&
  {Zucker}}]{Aschwanden1999}
{Aschwanden}, M.~J., {Newmark}, J.~S., {Delaboudini{\`e}re}, J.-P., {et~al.}
  1999, \bibinfo{title}{{Three-dimensional Stereoscopic Analysis of Solar
  Active Region Loops. I. SOHO/EIT Observations at Temperatures of (1.0-1.5)
  {\texttimes} {}10$^{6}$ K},} \apj, 515, 842, \dodoi{10.1086/307036}

\bibitem[{W.~T. {Barnes} {et~al.}(2019){Barnes}, {Bradshaw}, \&
  {Viall}}]{Barnes2019}
{Barnes}, W.~T., {Bradshaw}, S.~J., \& {Viall}, N.~M. 2019,
  \bibinfo{title}{{Understanding Heating in Active Region Cores through Machine
  Learning. I. Numerical Modeling and Predicted Observables},} \apj, 880, 56,
  \dodoi{10.3847/1538-4357/ab290c}

\bibitem[{W.~T. {Barnes} {et~al.}(2016){Barnes}, {Cargill}, \&
  {Bradshaw}}]{Barnes2016}
{Barnes}, W.~T., {Cargill}, P.~J., \& {Bradshaw}, S.~J. 2016,
  \bibinfo{title}{{Inference of Heating Properties from ``Hot'' Non-flaring
  Plasmas in Active Region Cores. I. Single Nanoflares},} \apj, 829, 31,
  \dodoi{10.3847/0004-637X/829/1/31}

\bibitem[{T. {Bastian} {et~al.}(2025{\natexlab{a}}){Bastian}, {Chen}, {Mondal},
  \& {Saint-Hilaire}}]{Bastian2025a}
{Bastian}, T., {Chen}, B., {Mondal}, S., \& {Saint-Hilaire}, P.
  2025{\natexlab{a}}, \bibinfo{title}{{Noise in Maps of the Sun at Radio
  Wavelengths I: Theoretical Considerations},} \solphys, 300, 91,
  \dodoi{10.1007/s11207-025-02499-9}

\bibitem[{T. {Bastian} {et~al.}(2025{\natexlab{b}}){Bastian}, {Chen}, {Mondal},
  \& {Saint-Hilaire}}]{Bastian2025b}
{Bastian}, T., {Chen}, B., {Mondal}, S., \& {Saint-Hilaire}, P.
  2025{\natexlab{b}}, \bibinfo{title}{{Noise in Maps of the Sun at Radio
  Wavelengths II: Solar Use Cases},} \solphys, 300, 90,
  \dodoi{10.1007/s11207-025-02498-w}

\bibitem[{M.~A. {Berger}(1991){Berger}}]{Berger1991}
{Berger}, M.~A. 1991, \bibinfo{title}{{Generation of coronal magnetic fields by
  random surface motions. I - Mean square twist and current density},} \aap,
  252, 369

\bibitem[{M.~A. {Berger}(1993){Berger}}]{Berger1993}
{Berger}, M.~A. 1993, \bibinfo{title}{{Energy-crossing number relations for
  braided magnetic fields},} \prl, 70, 705, \dodoi{10.1103/PhysRevLett.70.705}

\bibitem[{P.~K. {Browning} \& E.~R. {Priest}(1986){Browning} \&
  {Priest}}]{Browning1986}
{Browning}, P.~K., \& {Priest}, E.~R. 1986, \bibinfo{title}{{Heating of coronal
  arcades by magnetic tearing turbulence, using the Taylor-Heyvaerts
  hypothesis},} \aap, 159, 129

\bibitem[{P.~J. {Cargill}(2014){Cargill}}]{Cargill2014}
{Cargill}, P.~J. 2014, \bibinfo{title}{{Active Region Emission Measure
  Distributions and Implications for Nanoflare Heating},} \apj, 784, 49,
  \dodoi{10.1088/0004-637X/784/1/49}

\bibitem[{P.~J. {Cargill} {et~al.}(2012{\natexlab{a}}){Cargill}, {Bradshaw}, \&
  {Klimchuk}}]{Cargill2012a}
{Cargill}, P.~J., {Bradshaw}, S.~J., \& {Klimchuk}, J.~A. 2012{\natexlab{a}},
  \bibinfo{title}{{Enthalpy-based Thermal Evolution of Loops. II. Improvements
  to the Model},} \apj, 752, 161, \dodoi{10.1088/0004-637X/752/2/161}

\bibitem[{P.~J. {Cargill} {et~al.}(2012{\natexlab{b}}){Cargill}, {Bradshaw}, \&
  {Klimchuk}}]{Cargill2012b}
{Cargill}, P.~J., {Bradshaw}, S.~J., \& {Klimchuk}, J.~A. 2012{\natexlab{b}},
  \bibinfo{title}{{Enthalpy-based Thermal Evolution of Loops. III. Comparison
  of Zero-dimensional Models},} \apj, 758, 5, \dodoi{10.1088/0004-637X/758/1/5}

\bibitem[{J. {Chae} {et~al.}(2002){Chae}, {Poland}, \& {Aschwanden}}]{Chae2002}
{Chae}, J., {Poland}, A.~I., \& {Aschwanden}, M.~J. 2002,
  \bibinfo{title}{{Coronal Loops Heated by Magnetohydrodynamic Turbulence. I. A
  Model of Isobaric Quiet-Sun Loops with Constant Cross Sections},} \apj, 581,
  726, \dodoi{10.1086/344103}

\bibitem[{S.~R. {Cranmer} \& A.~R. {Winebarger}(2019){Cranmer} \&
  {Winebarger}}]{Cranmer2019}
{Cranmer}, S.~R., \& {Winebarger}, A.~R. 2019, \bibinfo{title}{{The Properties
  of the Solar Corona and Its Connection to the Solar Wind},} \araa, 57, 157,
  \dodoi{10.1146/annurev-astro-091918-104416}

\bibitem[{P. {Dmitruk} \& D.~O. {G{\'o}mez}(1997){Dmitruk} \&
  {G{\'o}mez}}]{Dmitruk1997}
{Dmitruk}, P., \& {G{\'o}mez}, D.~O. 1997, \bibinfo{title}{{Turbulent Coronal
  Heating and the Distribution of Nanoflares},} \apjl, 484, L83,
  \dodoi{10.1086/310760}

\bibitem[{P. {Dmitruk} \& D.~O. {G{\'o}mez}(1999){Dmitruk} \&
  {G{\'o}mez}}]{Dmitruk1999}
{Dmitruk}, P., \& {G{\'o}mez}, D.~O. 1999, \bibinfo{title}{{Scaling Law for the
  Heating of Solar Coronal Loops},} \apjl, 527, L63, \dodoi{10.1086/312390}

\bibitem[{G. {Einaudi} {et~al.}(1996){Einaudi}, {Velli}, {Politano}, \&
  {Pouquet}}]{Einaudi1996}
{Einaudi}, G., {Velli}, M., {Politano}, H., \& {Pouquet}, A. 1996,
  \bibinfo{title}{{Energy Release in a Turbulent Corona},} \apjl, 457, L113,
  \dodoi{10.1086/309893}

\bibitem[{G.~D. {Fleishman} {et~al.}(2021{\natexlab{a}}){Fleishman},
  {Anfinogentov}, {Stupishin}, {Kuznetsov}, \& {Nita}}]{Fleishman2021a}
{Fleishman}, G.~D., {Anfinogentov}, S.~A., {Stupishin}, A.~G., {Kuznetsov},
  A.~A., \& {Nita}, G.~M. 2021{\natexlab{a}}, \bibinfo{title}{{Coronal Heating
  Law Constrained by Microwave Gyroresonant Emission},} \apj, 909, 89,
  \dodoi{10.3847/1538-4357/abdab1}

\bibitem[{G.~D. {Fleishman} {et~al.}(2021{\natexlab{b}}){Fleishman},
  {Kuznetsov}, \& {Landi}}]{Fleishman2021b}
{Fleishman}, G.~D., {Kuznetsov}, A.~A., \& {Landi}, E. 2021{\natexlab{b}},
  \bibinfo{title}{{Gyroresonance and Free-Free Radio Emissions from
  Multithermal Multicomponent Plasma},} \apj, 914, 52,
  \dodoi{10.3847/1538-4357/abf92c}

\bibitem[{G.~D. {Fleishman} {et~al.}(2025){Fleishman}, {Kuznetsov}, \&
  {Nita}}]{Fleishman2025}
{Fleishman}, G.~D., {Kuznetsov}, A.~A., \& {Nita}, G.~M. 2025,
  \bibinfo{title}{{Steady-state Heating of Diffuse Coronal Plasma in a Solar
  Active Region},} \apj, 988, 100, \dodoi{10.3847/1538-4357/ade3dd}

\bibitem[{J.~M. {Fontenla} {et~al.}(2009){Fontenla}, {Curdt}, {Haberreiter},
  {Harder}, \& {Tian}}]{Fontenla2009}
{Fontenla}, J.~M., {Curdt}, W., {Haberreiter}, M., {Harder}, J., \& {Tian}, H.
  2009, \bibinfo{title}{{Semiempirical Models of the Solar Atmosphere. III. Set
  of Non-LTE Models for Far-Ultraviolet/Extreme-Ultraviolet Irradiance
  Computation},} \apj, 707, 482, \dodoi{10.1088/0004-637X/707/1/482}

\bibitem[{K. {Galsgaard} \& {\r{A}}. {Nordlund}(1996){Galsgaard} \&
  {Nordlund}}]{Galsgaard1996}
{Galsgaard}, K., \& {Nordlund}, {\r{A}}. 1996, \bibinfo{title}{{Heating and
  activity of the solar corona 1. Boundary shearing of an initially homogeneous
  magnetic field},} \jgr, 101, 13445, \dodoi{10.1029/96JA00428}

\bibitem[{K. {Galsgaard} \& {\r{A}}. {Nordlund}(1997){Galsgaard} \&
  {Nordlund}}]{Galsgaard1997}
{Galsgaard}, K., \& {Nordlund}, {\r{A}}. 1997, \bibinfo{title}{{Heating and
  activity of the solar corona. 2. Kink instability in a flux tube},} \jgr,
  102, 219, \dodoi{10.1029/96JA01462}

\bibitem[{L. {Golub} {et~al.}(1980){Golub}, {Maxson}, {Rosner}, {Vaiana}, \&
  {Serio}}]{Golub1980}
{Golub}, L., {Maxson}, C., {Rosner}, R., {Vaiana}, G.~S., \& {Serio}, S. 1980,
  \bibinfo{title}{{Magnetic fields and coronal heating},} \apj, 238, 343,
  \dodoi{10.1086/157990}

\bibitem[{G. {Halberstadt} \& J.~P. {Goedbloed}(1995){Halberstadt} \&
  {Goedbloed}}]{Halberstadt1995}
{Halberstadt}, G., \& {Goedbloed}, J.~P. 1995, \bibinfo{title}{{Alfven wave
  heating of coronal loops: photospheric excitation.},} \aap, 301, 559

\bibitem[{D.~L. {Hendrix} {et~al.}(1996){Hendrix}, {van Hoven}, {Mikic}, \&
  {Schnack}}]{Hendrix1996}
{Hendrix}, D.~L., {van Hoven}, G., {Mikic}, Z., \& {Schnack}, D.~D. 1996,
  \bibinfo{title}{{The Viability of Ohmic Dissipation as a Coronal Heating
  Source},} \apj, 470, 1192, \dodoi{10.1086/177942}

\bibitem[{J. {Heyvaerts} \& E.~R. {Priest}(1984){Heyvaerts} \&
  {Priest}}]{Heyvaerts1984}
{Heyvaerts}, J., \& {Priest}, E.~R. 1984, \bibinfo{title}{{Coronal heating by
  reconnection in DC current systems - A theory based on Taylor's hypothesis},}
  \aap, 137, 63

\bibitem[{J. {Heyvaerts} \& E.~R. {Priest}(1992){Heyvaerts} \&
  {Priest}}]{Heyvaerts1992}
{Heyvaerts}, J., \& {Priest}, E.~R. 1992, \bibinfo{title}{{A Self-consistent
  Turbulent Model for Solar Coronal Heating},} \apj, 390, 297,
  \dodoi{10.1086/171280}

\bibitem[{J.~V. {Hollweg}(1985){Hollweg}}]{Hollweg1985}
{Hollweg}, J.~V. 1985, in Advances in Space Plasma Physics, ed. W.~{Grossmann},
  E.~M. {Campbell}, \& B.~{Buti}, 77

\bibitem[{J.~V. {Hollweg}(1986){Hollweg}}]{Hollweg1986}
{Hollweg}, J.~V. 1986, \bibinfo{title}{{Transition region, corona, and solar
  wind in coronal holes},} \jgr, 91, 4111, \dodoi{10.1029/JA091iA04p04111}

\bibitem[{M. {Hossain} {et~al.}(1995){Hossain}, {Gray}, {Pontius}, {Matthaeus},
  \& {Oughton}}]{Hossain1995}
{Hossain}, M., {Gray}, P.~C., {Pontius}, Duane~H., J., {Matthaeus}, W.~H., \&
  {Oughton}, S. 1995, \bibinfo{title}{{Phenomenology for the decay of
  energy-containing eddies in homogeneous MHD turbulence},} Physics of Fluids,
  7, 2886, \dodoi{10.1063/1.868665}

\bibitem[{G.~W. {Inverarity} \& E.~R. {Priest}(1995{\natexlab{a}}){Inverarity}
  \& {Priest}}]{Inverarity1995a}
{Inverarity}, G.~W., \& {Priest}, E.~R. 1995{\natexlab{a}},
  \bibinfo{title}{{Turbulent coronal heating. II. Twisted flux tube.},} \aap,
  296, 395

\bibitem[{G.~W. {Inverarity} \& E.~R. {Priest}(1995{\natexlab{b}}){Inverarity}
  \& {Priest}}]{Inverarity1995b}
{Inverarity}, G.~W., \& {Priest}, E.~R. 1995{\natexlab{b}},
  \bibinfo{title}{{Turbulent coronal heating. III. Wave heating in coronal
  loops.},} \aap, 302, 567

\bibitem[{G.~W. {Inverarity} {et~al.}(1995){Inverarity}, {Priest}, \&
  {Heyvaerts}}]{Inverarity1995}
{Inverarity}, G.~W., {Priest}, E.~R., \& {Heyvaerts}, J. 1995,
  \bibinfo{title}{{Turbulent coronal heating. I. Sheared arcade.},} \aap, 293,
  913

\bibitem[{R. {Jain} \& C.~H. {Mandrini}(2006){Jain} \& {Mandrini}}]{Jain2006}
{Jain}, R., \& {Mandrini}, C.~H. 2006, \bibinfo{title}{{The relationship
  between magnetic field strength and loop lengths in solar coronal active
  regions},} \aap, 450, 375, \dodoi{10.1051/0004-6361:20053619}

\bibitem[{J.~A. {Klimchuk}(2015){Klimchuk}}]{Klimchuk2015}
{Klimchuk}, J.~A. 2015, \bibinfo{title}{{Key aspects of coronal heating},}
  Philosophical Transactions of the Royal Society of London Series A, 373,
  20140256, \dodoi{10.1098/rsta.2014.0256}

\bibitem[{J.~A. {Klimchuk} {et~al.}(2008){Klimchuk}, {Patsourakos}, \&
  {Cargill}}]{Klimchuk2008}
{Klimchuk}, J.~A., {Patsourakos}, S., \& {Cargill}, P.~J. 2008,
  \bibinfo{title}{{Highly Efficient Modeling of Dynamic Coronal Loops},} \apj,
  682, 1351, \dodoi{10.1086/589426}

\bibitem[{J.~A. {Klimchuk} \& L.~J. {Porter}(1995){Klimchuk} \&
  {Porter}}]{Klimchuk1995}
{Klimchuk}, J.~A., \& {Porter}, L.~J. 1995, \bibinfo{title}{{Scaling of heating
  rates in solar coronal loops},} \nat, 377, 131, \dodoi{10.1038/377131a0}

\bibitem[{A. {Kuznetsov}(2025{\natexlab{a}}){Kuznetsov}}]{Kuznetsov2025a}
{Kuznetsov}, A. 2025{\natexlab{a}}, \bibinfo{title}{{Rendering Codes for Solar
  Coronal Microwave and EUV Emissions},}, 1.0.0 Zenodo,
  \dodoi{10.5281/zenodo.15389253}

\bibitem[{A. {Kuznetsov}(2025{\natexlab{b}}){Kuznetsov}}]{Kuznetsov2025b}
{Kuznetsov}, A. 2025{\natexlab{b}}, \bibinfo{title}{{Coronal Heating Model
  Pipeline},}, 1.0.0 Zenodo, \dodoi{10.5281/zenodo.15373435}

\bibitem[{A. {Kuznetsov} {et~al.}(2021){Kuznetsov}, {Fleishman}, \&
  {Landi}}]{Kuznetsov2021}
{Kuznetsov}, A., {Fleishman}, G., \& {Landi}, E. 2021, \bibinfo{title}{{Codes
  for computing the solar gyroresonance and free-free radio emissions: the
  first release},}, v1.0.0 Zenodo, \dodoi{10.5281/zenodo.4625572}

\bibitem[{A. {Kuznetsov} {et~al.}(2022){Kuznetsov}, {Nita}, {Fleishman}, \&
  {Anfinogentov}}]{Kuznetsov2022}
{Kuznetsov}, A., {Nita}, G., {Fleishman}, G., \& {Anfinogentov}, S. 2022, in
  44th COSPAR Scientific Assembly. Held 16-24 July, Vol.~44, 1219

\bibitem[{J. {Lee}(2007){Lee}}]{Lee2007}
{Lee}, J. 2007, \bibinfo{title}{{Radio Emissions from Solar Active Regions},}
  \ssr, 133, 73, \dodoi{10.1007/s11214-007-9206-2}

\bibitem[{C. {Mac Cormack} {et~al.}(2020){Mac Cormack}, {L{\'o}pez Fuentes},
  {Mandrini}, {Lloveras}, {Poisson}, \& {V{\'a}squez}}]{MacCormack2020}
{Mac Cormack}, C., {L{\'o}pez Fuentes}, M., {Mandrini}, C.~H., {et~al.} 2020,
  \bibinfo{title}{{Scaling laws of quiet-Sun coronal loops},} Advances in Space
  Research, 65, 1616, \dodoi{10.1016/j.asr.2019.08.019}

\bibitem[{C.~H. {Mandrini} {et~al.}(2000){Mandrini}, {D{\'e}moulin}, \&
  {Klimchuk}}]{Mandrini2000}
{Mandrini}, C.~H., {D{\'e}moulin}, P., \& {Klimchuk}, J.~A. 2000,
  \bibinfo{title}{{Magnetic Field and Plasma Scaling Laws: Their Implications
  for Coronal Heating Models},} \apj, 530, 999, \dodoi{10.1086/308398}

\bibitem[{L.~J. {Milano} {et~al.}(1997){Milano}, {G{\'o}mez}, \&
  {Martens}}]{Milano1997}
{Milano}, L.~J., {G{\'o}mez}, D.~O., \& {Martens}, P. C.~H. 1997,
  \bibinfo{title}{{Solar Coronal Heating: AC versus DC},} \apj, 490, 442,
  \dodoi{10.1086/304845}

\bibitem[{B. {Mondal} {et~al.}(2024){Mondal}, {Athiray}, {Winebarger},
  {Savage}, {Kobayashi}, {Bradshaw}, {Barnes}, {Champey}, {Cheimets},
  {Dud{\'\i}k}, {Golub}, {Mason}, {McKenzie}, {Moore}, {Madsen}, {Reeves},
  {Testa}, {Vigil}, {Warren}, {Walsh}, \& {Del Zanna}}]{Mondal2024}
{Mondal}, B., {Athiray}, P.~S., {Winebarger}, A.~R., {et~al.} 2024,
  \bibinfo{title}{{Determining the Nanoflare Heating Frequency of an X-Ray
  Bright Point Observed by MaGIXS},} \apj, 967, 23,
  \dodoi{10.3847/1538-4357/ad2766}

\bibitem[{G.~M. {Nita} {et~al.}(2018){Nita}, {Viall}, {Klimchuk},
  {Loukitcheva}, {Gary}, {Kuznetsov}, \& {Fleishman}}]{Nita2018}
{Nita}, G.~M., {Viall}, N.~M., {Klimchuk}, J.~A., {et~al.} 2018,
  \bibinfo{title}{{Dressing the Coronal Magnetic Extrapolations of Active
  Regions with a Parameterized Thermal Structure},} \apj, 853, 66,
  \dodoi{10.3847/1538-4357/aaa4bf}

\bibitem[{G.~M. {Nita} {et~al.}(2023){Nita}, {Fleishman}, {Kuznetsov},
  {Anfinogentov}, {Stupishin}, {Kontar}, {Schonfeld}, {Klimchuk}, \&
  {Gary}}]{Nita2023}
{Nita}, G.~M., {Fleishman}, G.~D., {Kuznetsov}, A.~A., {et~al.} 2023,
  \bibinfo{title}{{Data-constrained Solar Modeling with GX Simulator},} \apjs,
  267, 6, \dodoi{10.3847/1538-4365/acd343}

\bibitem[{L. {Ofman} {et~al.}(1995){Ofman}, {Davila}, \&
  {Steinolfson}}]{Ofman1995}
{Ofman}, L., {Davila}, J.~M., \& {Steinolfson}, R.~S. 1995,
  \bibinfo{title}{{Coronal Heating by the Resonant Absorption of Alfven Waves:
  Wavenumber Scaling Laws},} \apj, 444, 471, \dodoi{10.1086/175621}

\bibitem[{D.~E. {Osterbrock}(1961){Osterbrock}}]{Osterbrock1961}
{Osterbrock}, D.~E. 1961, \bibinfo{title}{{The Heating of the Solar
  Chromosphere, Plages, and Corona by Magnetohydrodynamic Waves.},} \apj, 134,
  347, \dodoi{10.1086/147165}

\bibitem[{E.~N. {Parker}(1983){Parker}}]{Parker1983}
{Parker}, E.~N. 1983, \bibinfo{title}{{Magnetic Neutral Sheets in Evolving
  Fields - Part Two - Formation of the Solar Corona},} \apj, 264, 642,
  \dodoi{10.1086/160637}

\bibitem[{E.~N. {Parker}(1988){Parker}}]{Parker1988}
{Parker}, E.~N. 1988, \bibinfo{title}{{Nanoflares and the Solar X-Ray Corona},}
  \apj, 330, 474, \dodoi{10.1086/166485}

\bibitem[{E.~R. {Priest} {et~al.}(2018){Priest}, {Chitta}, \&
  {Syntelis}}]{Priest2018}
{Priest}, E.~R., {Chitta}, L.~P., \& {Syntelis}, P. 2018, \bibinfo{title}{{A
  Cancellation Nanoflare Model for Solar Chromospheric and Coronal Heating},}
  \apjl, 862, L24, \dodoi{10.3847/2041-8213/aad4fc}

\bibitem[{B. {Roberts}(2000){Roberts}}]{Roberts2000}
{Roberts}, B. 2000, \bibinfo{title}{{Waves and Oscillations in the Corona -
  (Invited Review)},} \solphys, 193, 139, \dodoi{10.1023/A:1005237109398}

\bibitem[{M.~S. {Ruderman} {et~al.}(1997){Ruderman}, {Berghmans}, {Goossens},
  \& {Poedts}}]{Ruderman1997}
{Ruderman}, M.~S., {Berghmans}, D., {Goossens}, M., \& {Poedts}, S. 1997,
  \bibinfo{title}{{Direct excitation of resonant torsional Alfven waves by
  footpoint motions.},} \aap, 320, 305

\bibitem[{P.~H. {Scherrer} {et~al.}(2012){Scherrer}, {Schou}, {Bush},
  {Kosovichev}, {Bogart}, {Hoeksema}, {Liu}, {Duvall}, {Zhao}, {Title},
  {Schrijver}, {Tarbell}, \& {Tomczyk}}]{Scherrer2012}
{Scherrer}, P.~H., {Schou}, J., {Bush}, R.~I., {et~al.} 2012,
  \bibinfo{title}{{The Helioseismic and Magnetic Imager (HMI) Investigation for
  the Solar Dynamics Observatory (SDO)},} \solphys, 275, 207,
  \dodoi{10.1007/s11207-011-9834-2}

\bibitem[{S.~J. {Schonfeld}(2022){Schonfeld}}]{Schonfeld2022}
{Schonfeld}, S.~J. 2022, \bibinfo{title}{{schonfsj/ebtel\_parallel:
  ebtel\_parallel v1.0.0},}, v1.0.0 Zenodo, \dodoi{10.5281/zenodo.7154827}

\bibitem[{S.~J. Schonfeld \& J.~A. Klimchuk(2020)Schonfeld \&
  Klimchuk}]{Schonfeld2020}
Schonfeld, S.~J., \& Klimchuk, J.~A. 2020, \bibinfo{title}{{Transition Region
  Contribution to AIA Observations in the Context of Coronal Heating},} The
  Astrophysical Journal, 905, 12, \dodoi{10.3847/1538-4357/abc3bd}

\bibitem[{P.~A. {Sturrock} \& Y. {Uchida}(1981){Sturrock} \&
  {Uchida}}]{Sturrock1981}
{Sturrock}, P.~A., \& {Uchida}, Y. 1981, \bibinfo{title}{{Coronal heating by
  stochastic magnetic pumping},} \apj, 246, 331, \dodoi{10.1086/158926}

\bibitem[{I. {Ugarte-Urra} {et~al.}(2019){Ugarte-Urra}, {Crump}, {Warren}, \&
  {Wiegelmann}}]{UgarteUrra2019}
{Ugarte-Urra}, I., {Crump}, N.~A., {Warren}, H.~P., \& {Wiegelmann}, T. 2019,
  \bibinfo{title}{{The Magnetic Properties of Heating Events on
  High-temperature Active-region Loops},} \apj, 877, 129,
  \dodoi{10.3847/1538-4357/ab1d4d}

\bibitem[{A.~A. {van Ballegooijen}(1986){van
  Ballegooijen}}]{vanBallegooijen1986}
{van Ballegooijen}, A.~A. 1986, \bibinfo{title}{{Cascade of Magnetic Energy as
  a Mechanism of Coronal Heating},} \apj, 311, 1001, \dodoi{10.1086/164837}

\bibitem[{A.~A. {van Ballegooijen} \& S.~R. {Cranmer}(2008){van Ballegooijen}
  \& {Cranmer}}]{vanBallegooijen2008}
{van Ballegooijen}, A.~A., \& {Cranmer}, S.~R. 2008,
  \bibinfo{title}{{Hyperdiffusion as a Mechanism for Solar Coronal Heating},}
  \apj, 682, 644, \dodoi{10.1086/587457}

\bibitem[{G.~E. {Vekstein} {et~al.}(1993){Vekstein}, {Priest}, \&
  {Steele}}]{Vekstein1993}
{Vekstein}, G.~E., {Priest}, E.~R., \& {Steele}, C.~D.~C. 1993,
  \bibinfo{title}{{On the Problem of Magnetic Coronal Heating by Turbulent
  Relaxation},} \apj, 417, 781, \dodoi{10.1086/173358}

\bibitem[{N.~M. {Viall} \& J.~A. {Klimchuk}(2017){Viall} \&
  {Klimchuk}}]{Viall2017}
{Viall}, N.~M., \& {Klimchuk}, J.~A. 2017, \bibinfo{title}{{A Survey of
  Nanoflare Properties in Active Regions Observed with the Solar Dynamics
  Observatory},} \apj, 842, 108, \dodoi{10.3847/1538-4357/aa7137}

\bibitem[{G.~L. {Withbroe} \& R.~W. {Noyes}(1977){Withbroe} \&
  {Noyes}}]{Withbroe1977}
{Withbroe}, G.~L., \& {Noyes}, R.~W. 1977, \bibinfo{title}{{Mass and energy
  flow in the solar chromosphere and corona.},} \araa, 15, 363,
  \dodoi{10.1146/annurev.aa.15.090177.002051}

\bibitem[{K.~E. {Yang} {et~al.}(2018){Yang}, {Longcope}, {Ding}, \&
  {Guo}}]{Yang2018}
{Yang}, K.~E., {Longcope}, D.~W., {Ding}, M.~D., \& {Guo}, Y. 2018,
  \bibinfo{title}{{Observationally quantified reconnection providing a viable
  mechanism for active region coronal heating},} Nature Communications, 9, 692,
  \dodoi{10.1038/s41467-018-03056-8}

\bibitem[{Y. {Zhou} \& W.~H. {Matthaeus}(1990){Zhou} \& {Matthaeus}}]{Zhou1990}
{Zhou}, Y., \& {Matthaeus}, W.~H. 1990, \bibinfo{title}{{Models of inertial
  range spectra of interplanetary magnetohydrodynamic turbulence},} \jgr, 95,
  14881, \dodoi{10.1029/JA095iA09p14881}

\bibitem[{H. {Zirin} {et~al.}(1991){Zirin}, {Baumert}, \&
  {Hurford}}]{Zirin1991}
{Zirin}, H., {Baumert}, B.~M., \& {Hurford}, G.~J. 1991, \bibinfo{title}{{The
  Microwave Brightness Temperature Spectrum of the Quiet Sun},} \apj, 370, 779,
  \dodoi{10.1086/169861}

\end{thebibliography}

\appendix
\section{Comparing the observed and synthetic images}\label{MultiImageComparison}
\begin{figure*}
\centerline{\includegraphics{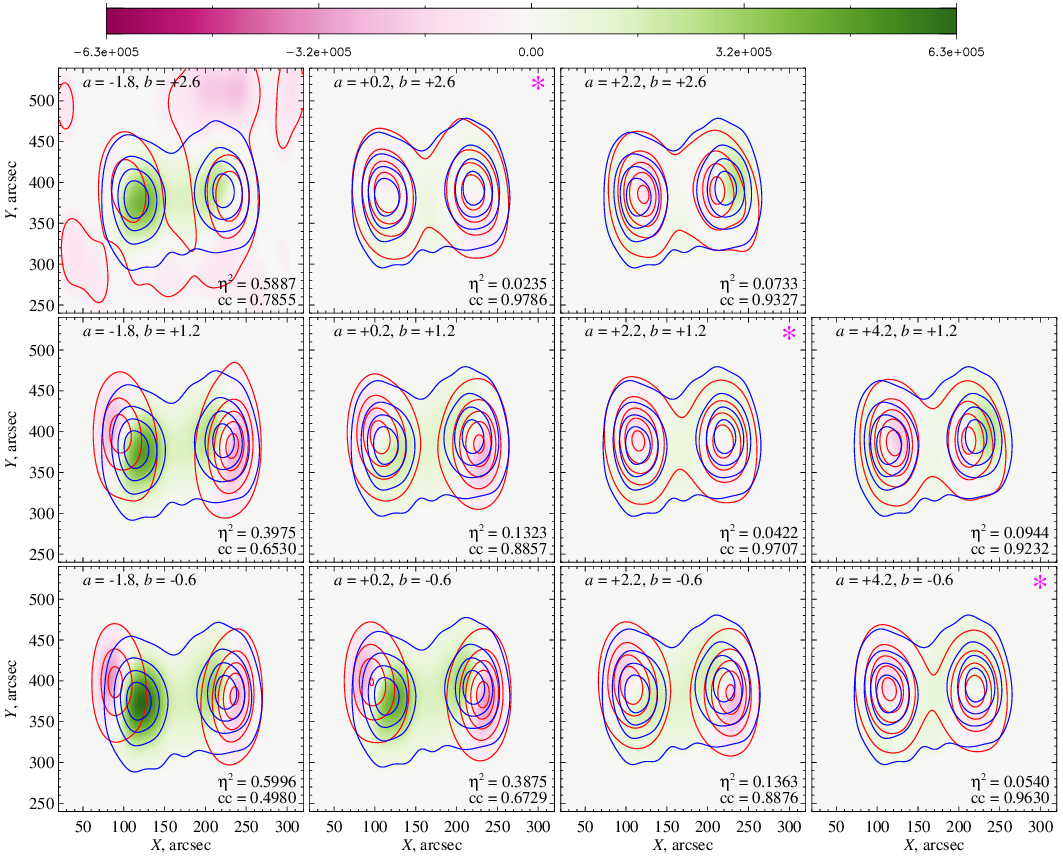}}
\caption{Comparison of the observed and synthetic (model) microwave images of AR 12396, at 2022-01-31 04:20 UT, at the frequency of 5.80 GHz. The synthetic images were computed for the steady-state EBTEL heating regime, for chosen $a$ and $b$ values (shown in the panels), and for $Q_0$ values providing the minimum $\eta^2$ metrics. The color background shows the observations-minus-model residuals (displayed only in the areas where either observed or model fluxes exceed 10\% of their respective maxima). The red and blue contours represent respectively the model and observed fluxes (brightness temperatures) at the levels of 0.06, 0.19, 0.32, 0.44, and 0.57 MK. The panels with the $(a, b)$ combinations located within the ``degeneracy stripe'' are marked by magenta asterisks. The panel in the upper-right corner is missing because no fit for $(a, b)=(4.2, 2.6)$ was obtained due to the limited EBTEL output table.}
\label{MultiABcomparison}
\end{figure*}

In this Section, we compare the synthetic microwave images for different parameters of the coronal heating model, including those that provide ``good'' and ``not-so-good'' agreement with observations. For an example case (for AR 12936 at 2022-01-31 04:20 UT, at the frequency of 5.80 GHz, for the steady-state EBTEL heating regime), we have selected a grid of representative $(a, b)$ combinations---at $a=-1.8$, 0.2, 2.2, and 4.2, and $b=-0.6$, 1.2, and 2.6; some of these points are shown in the upper-left panel of Figure \ref{FigMultiEBTEL}. Figure \ref{MultiABcomparison} demonstrates the corresponding observations-minus-model residual maps, $\eta^2$ metrics and correlation coefficients; the synthetic and observed microwave images are shown by overplotted contours at the same brightness temperature levels.

The $(a, b)$ combinations that provide the lowest $\eta^2$ values (e.g., $a=0.2$ and $b=2.6$, and, to some degree, two other combinations located within the ``degeneracy stripe'') also provide the highest correlation coefficients, a good match between the model and observed intensity contours, and the lowest residuals (although the residuals still exceed the estimated measurement uncertainties, see $2^{\mathrm{nd}}$ paragraph of Section \ref{Results}). In contrast, when the $(a, b)$ parameters deviate considerably from their best-fit values (like, e.g., in the panels for $a=-1.8$ and $b=-0.6$), the model and observed intensity contours do not match each other, which results in larger residuals and $\eta^2$ metrics and lower correlation coefficients; i.e., such heating models, although able to reproduce the total microwave flux, fail to reproduce adequately the emission source structure (including the distance between the two sunspot-associated source components and the intensity ratio of those components). Therefore we conclude that a) the $\eta^2$ metric (\ref{EtaMetric}) is a reliable parameter to quantify the model-to-observations agreement, b) variations of the heating model parameters within the considered range result in variations of the observations-minus-model residuals well exceeding the measurement uncertainty in the microwave images, and hence c) the used algorithm indeed allows us to compare different coronal heating models (characterized by the $a$ and $b$ indices and the EBTEL lookup table) and select the one(s) that reflect the actual heating processes most closely.

\section{Fitting the degeneracy stripes}\label{StripeFitting}
Fitting the degeneracy stripes in $a{-}b$ diagrams with straight lines was performed within certain areas determined by the condition $\eta^2<k\eta^2_{\min}$, typically with $k=2$. The fitting procedure included the following steps:

a) For each value of $a$, find the value of $b$ that provides the minimum fitting quality metric in the corresponding column of the diagram. Fit the obtained set of points $(a_i, b_i)$, $i=1, \ldots, N_a$, with a straight line in the form of $b=p+qa$, and estimate the uncertainties of the fit parameters $\sigma_p$ and $\sigma_q$.

b) For each value of $b$, find the value of $a$ that provides the minimum fitting quality metric in the corresponding row of the diagram. Fit the obtained set of points $(b_i, a_i)$, $i=1, \ldots, N_b$, with a straight line in the form of $a=\tilde p+\tilde qb$, and estimate the uncertainties of the fit parameters $\tilde\sigma_{p}$ and $\tilde\sigma_q$.

c) The resulting (``average'') fit is a straight line that represents a bisector of the acute angle formed by the two straight lines obtained at previous steps, and passes through the point of their intersection. Its equation has the form of $b=b_0+a\delta$, where
\begin{displaymath}
\delta=\tan\left(\frac{1}{2}\arctan q+\frac{1}{2}\operatorname{arccot}\tilde q\right),
\end{displaymath}
\begin{displaymath}
b_0=\frac{p+\tilde pq-\delta(\tilde p+p\tilde q)}{1-q\tilde q}.
\end{displaymath}
The uncertainty of the slope $\delta$ is given by
\begin{displaymath}
\sigma_{\delta}=\frac{1+\delta^2}{2}\left(\frac{\sigma_q}{1+q^2}+\frac{\tilde\sigma_q}{1+{\tilde q}^2}\right).
\end{displaymath}
The uncertainty of the intercept $b_0$ is given by a more complicated expression; it is more convenient to estimate this value numerically. The presented algorithm has been found to provide a reasonably good visual agreement with the features in the $a{-}b$ diagrams.

The distance from a straight line $b=b_0+a\delta$ to a given point $(a_*, b_*)$ is given by
\begin{displaymath}
d=\frac{|b_0+a_*\delta-b_*|}{\sqrt{1+\delta^2}}.
\end{displaymath}
Therefore, width of a degeneracy stripe $w$ at a certain level (determined relatively to $\eta^2_{\min}$) can be estimated as
\begin{displaymath}
w=\max(d_{\mathrm{above}})+\max(d_{\mathrm{below}}),
\end{displaymath}
where $d_{\mathrm{above}}$ and $d_{\mathrm{below}}$ are the distances to the points located in the $a{-}b$ diagram respectively above and below the best-fit straight line of the stripe, within the area where $\eta^2<k\eta^2_{\min}$.

\section{Estimating the cooling times}\label{CoolingTimeFormulae}
Following \citet{Cargill2014}, for the radiative loss function in the form of $\Lambda(T)=\chi T^{\alpha}$, the effective (i.e., determined by both the conductive and radiative heat losses) plasma cooling time $\tau_{\mathrm{c}}$ in a coronal magnetic loop can be estimated as
\begin{displaymath}
\tau_{\mathrm{c}}=3k_{\mathrm{B}}\left(\frac{2-\alpha}{1-\alpha}\right)\left[\frac{1}{\kappa_0^{4-2\alpha}\chi^7}\frac{L^{8-4\alpha}}{(n_0T_0)^{3+2\alpha}}\right]^{1/(11-2\alpha)},
\end{displaymath}
where $L$ is the loop length, $n_0$ and $T_0$ are the initial plasma density and temperature, $\kappa_0=1.0\times 10^{-6}$ erg $\textrm{cm}^{-1}$ $\textrm{s}^{-1}$ $\textrm{K}^{-7/2}$ is the Spitzer coefficient of thermal conductivity, and $k_{\mathrm{B}}$ is the Boltzmann constant.

In this study, we use the radiative loss function in the form of \citep{Klimchuk2008}
\begin{displaymath}
\Lambda(T)=\left\{\begin{array}{ll}
1.09\times 10^{-31}T^2, & T\le 10^{4.97},\\
8.87\times 10^{-17}T^{-1}, & 10^{4.97}<T\le 10^{5.67},\\
1.90\times 10^{-22}, & 10^{5.67}<T\le 10^{6.18},\\
3.53\times 10^{-13}T^{-3/2}, & 10^{6.18}<T\le 10^{6.55},\\
3.46\times 10^{-25}T^{1/3}, & 10^{6.55}<T\le 10^{6.90},\\
5.49\times 10^{-16}T^{-1}, & 10^{6.90}<T\le 10^{7.63},\\
1.96\times 10^{-27}T^{1/2}, & 10^{7.63}<T.
\end{array}\right.
\end{displaymath}
Here, the radiative loss function $\Lambda(T)$ is measured in units of erg $\textrm{cm}^3$ $\textrm{s}^{-1}$.
\end{document}